\documentclass[aps,prx,reprint,superscriptaddress,amsmath,amssymb,floatfix,nofootinbib,longbibliography]{revtex4-2}

\usepackage[T1]{fontenc}
\usepackage[utf8]{inputenc}
\usepackage{lmodern}
\usepackage{microtype}
\usepackage{amsmath,amssymb,mathtools,bm}
\usepackage{graphicx}
\usepackage{booktabs}
\usepackage{xcolor}
\usepackage[hidelinks]{hyperref}
\usepackage[nameinlink,noabbrev]{cleveref}
\makeatletter
\expandafter\let\expandafter\widefigure\csname figure*\endcsname
\expandafter\let\expandafter\endwidefigure\csname endfigure*\endcsname
\RenewDocumentEnvironment{figure}{O{!t}}{\widefigure[!t]}{\endwidefigure}
\RenewDocumentEnvironment{figure*}{O{!t}}{\widefigure[!t]}{\endwidefigure}
\makeatother
\crefname{figure}{Fig.}{Figs.}

\hypersetup{
  colorlinks=true,
  linkcolor=blue!45!black,
  citecolor=blue!45!black,
  urlcolor=blue!45!black,
  pdftitle={Hydrodynamics of perceptual matter: from neural representations to collective motion},
  pdfauthor={Mohammad Salahshour},
  pdfsubject={Kinetic and hydrodynamic theory linking neural representations to collective motion in perceptual matter},
  pdfkeywords={perceptual matter, Harmonic Theory, active matter, collective motion, bearing interactions, kinetic theory}
}
\allowdisplaybreaks

\newcommand{\ee}{\bm e}
\newcommand{\rr}{\bm r}
\newcommand{\xx}{\bm x}
\newcommand{\dd}{\mathrm d}
\newcommand{\ii}{\mathrm i}
\newcommand{\e}{\mathrm e}
\newcommand{\order}{\mathcal O}
\newcommand{\Jcal}{\mathcal J}
\newcommand{\Dr}{D_{\mathrm r}}
\newcommand{\meanrho}{\bar\rho}
\crefname{equation}{Eq.}{Eqs.}
\begin{document}
\title{
Hydrodynamics of perceptual matter:\\
from neural representations to collective motion}

\author{Mohammad Salahshour}

\email[{}]{salahshour.mohammad@gmail.com}

\affiliation{Department of Collective Behaviour, Max Planck Institute of Animal Behavior, Konstanz, Germany}
\affiliation{Centre for the Advanced Study of Collective Behaviour, University of Konstanz, Konstanz, Germany}
\affiliation{Department of Biology, University of Konstanz, Konstanz, Germany}

\begin{abstract}

How do neural representations of neighbors become collective material
behavior?  Harmonic Theory (HT) reduces bearing representations to signed
perceptual channels traceable to sensory acuity, decision dynamics, and interaction
distance.  Here we derive a nonlocal kinetic and hydrodynamic theory from the HT
particle law without introducing heading alignment.  The resulting fields retain HT's order-resolved perceptual spectrum and keep the neural and sensory origin of each coefficient visible.
The theory predicts a torque-driven density instability: HT's direct-approach
channel turns agents toward density excesses, while self-propulsion carries them
inward.  For isotropic spatial interactions and positive rotational diffusion, perceptual channel \(n\) generically first
affects conserved density at spatial order \(k^{2n}\), and the full radial transforms
select finite pattern wavelengths.  Two populations can share the same long-wave instability threshold yet select different initial pattern spacings because their higher perceptual channels differ.
Beyond onset, the one-particle continuum sustains a bidirectional nematic column;
a prescribed, symmetry-allowed two-stream response amplifies a directional imbalance into sustained polar flow.  In a distinct pure-repulsion regime,
one-particle calculations show polar transport and translating transverse
bands, with late pattern selection sensitive to numerical transport.  The framework connects neural and sensory mechanisms to
continuum coefficients, instabilities, and alignment-free collective motion.

\end{abstract}
\maketitle

\section{Introduction}
Models with prescribed interaction rules helped establish the modern theory of collective motion.  Reynolds'
separation, alignment, and cohesion rules~\citep{Reynolds1987}, the Vicsek
alignment transition~\citep{Vicsek1995}, and spatial-zone models of attraction and
repulsion~\citep{Couzin2002} showed how local responses can organize a group.  Their
success does not, by itself, identify the biological mechanism behind a rule~\citep{Sayin2025}. Different interaction laws can be compatible with similar group-level
patterns~\citep{Katz2011,HerbertRead2011}, and a prescribed turning rule does not
explain how a nervous system represents a scene and converts it into a movement
decision~\citep{GomezMarin2014,Krakauer2017}.

Collective motion can arise from visual interactions without explicit
heading alignment. Attraction restricted to a vision cone produces moving files
and nematic structures~\citep{BarberisPeruani2016}, while responses to a visual
projection field generate coordinated motion without comparing neighbours'
headings~\citep{BastienRomanczuk2020}.
\emph{Allocentric Flocking} (AF) develops the connection between perception and collective behaviour at the level of internal neural dynamics.  Neighbour
bearings are projected onto an internal directional representation, in which a localized decision profile encodes the selected movement direction. Coordinated motion then arises without comparing
neighbour headings~\citep{SalahshourCouzin2025}.  The essential step is the
projection of sensory input onto a directional manifold.  In the present
two-dimensional setting, this internal representation is a circle.  Neural heading systems
provide biological precedents for such representations
~\citep{Zhang1996,SeeligJayaraman2015,Kim2017,TurnerEvans2017}.

\emph{Harmonic Theory of Behavior} (HT) explains why this construction
is more general than one neural implementation~\citep{SalahshourCouzin2026HT}.
HT separates the internal dynamics that maintain a directional decision state from
the environmental coupling through which sensory input changes that state.  The
internal mechanism need not be AF's ring attractor~\citep{SalahshourCouzin2025}. Any sufficiently fast,
rotation-symmetric dynamics that support a localized decision state can play the
same role~\citep{SalahshourCouzin2026HT}.  Coarse-graining the neural and sensory dynamics then produces signed
perceptual interaction channels at the particle level.  Each channel retains three
distinct origins---how interaction strength varies with distance (the HT radial filter), how accurately a
neighbour is sensed (the HT sensory filter), and how sensory input is converted into a directional
decision (the HT decision filter).  This spectral structure also enables harmonic spectroscopy of animal decisions, in which cue arrangements selectively cancel or preserve perceptual channels to measure their contributions~\citep{Salahshour2026Spectroscopy}. HT therefore connects neural mechanism to individual motion,
but it stops before describing the group as a continuous material. Here, we fill this gap.
{}

Most continuum theories of active matter begin at a later descriptive
level, either with macroscopic fields chosen from symmetry, or with an already
specified particle rule~\citep{TonerTu1995,Marchetti2013,Bertin2006,Peshkov2014}.
Collective motion without explicit heading alignment is also not unique to AF or
HT.  At the particle and experimental levels, position- and vision-based responses
can organize moving groups~\citep{BastienRomanczuk2020,Lavergne2019}.  At the
continuum level, vision-cone attraction can produce aggregation, nematic bands, and
moving locally polar structures~\citep{BarberisPeruani2016,Peruani2017}. Turning toward dense regions can drive phase separation
~\citep{ZhangAlertYanWingreenGranick2021}. Pair correlations can also produce polar
motion from turn-away interactions~\citep{DasCiarchiZhouYanZhangAlert2024}.  Building on these alignment-free continuum theories, we derive hydrodynamics from a mechanistic description of neural sensory processing, completing the coarse-graining that HT begins.

Methodologically, we carry HT's signed
perceptual channels from interacting agents into continuous density and heading
fields while retaining their explicit sensory, decision, and radial factors.  The
resulting transport and instability parameters can therefore be traced back to
sensory acuity, directional decision-making, distance-dependent interaction
strength, and ultimately the internal neural dynamics represented in HT~\citep{SalahshourCouzin2026HT}.  The continuum theory thus connects the mechanisms of perception to the dynamics of the
material formed by many such agents.  We call it the hydrodynamics of perceptual matter.

Preserving this microscopic structure also yields distinct
physical predictions.  Inward turning toward a density excess is known to create
an instability~\citep{ZhangAlertYanWingreenGranick2021}. Here we show that its threshold separates into the same
sensory, decision, and distance factors that HT assigns to direct approach by one
individual.  Successive perceptual channels first influence density at
successively higher spatial-derivative orders, while the full distance dependence
determines which finite wavelength grows.
Square and Gaussian decision profiles can share the first perceptual channel and
instability threshold yet select different initial pattern spacings. Matching
their first two channels nearly recovers the same growth curve.  Beyond onset, the one-particle
continuum sustains an attractive, bidirectional nematic column but does not choose
between its two travel directions during the simulated interval.  A prescribed, symmetry-allowed two-stream response amplifies a small directional
imbalance into delayed polar flow, illustrating a possible role for the pair
information omitted by the one-particle description.  Pure-repulsion calculations also sustain collective currents without a pair
term, while their density patterns remain sensitive to numerical transport.
These predictions are not supplied by
generic alignment-free hydrodynamics without adding HT's perceptual decomposition
~\citep{Peruani2017,ZhangAlertYanWingreenGranick2021,DasCiarchiZhouYanZhangAlert2024}.
Together they establish a hydrodynamics of perceptual matter in which both
macroscopic coefficients and collective phenomena remain connected to how
individuals represent and act on sensory information.

\section{From perceptual agents to hydrodynamics}
\label{sec:math-roadmap}

This section develops the central theory.  We begin with HT's microscopic perceptual law, derive its
nonlocal kinetic continuum, resolve the heading distribution into fields, and
then obtain the hydrodynamic density equation.  

\begingroup

Detailed derivations, validity conditions, numerical protocols,
supporting figures, and supplemental video captions are provided in
the Supplemental Material (SM)~\citep{SupplementalMaterial}.
\par\endgroup

Because the derivation uses three Fourier decompositions, we fix the
terminology first.  An~\emph{HT perceptual harmonic}~\citep{SalahshourCouzin2026HT} of order
\(n\) means the bearing-response coefficient \(q_n\) in
Eq.~\eqref{eq:coefficients-methods} and its continuum cue \(C_n[\rho]\) in
Eq.~\eqref{eq:coarse-cue-main}.  The fields \(f_m\) in
Eq.~\eqref{eq:angular-transform-methods} are \emph{heading moments} of the kinetic
distribution, indexed by \(m\).  Density variations labelled by a wavevector
\(\bm k\), including variations at \(2\bm k\), are \emph{spatial Fourier modes}.
An eigenvector of the linearized dynamics is a \emph{linear eigenmode}. The prescribed ordered-state population response is a \emph{two-stream response}. We use
``harmonic'' alone only as part of the name Harmonic Theory.

\subsection{From HT's bearing law to perceptual fields}
\label{sec:perceptual-fields}
\label{sec:methods-particle}

HT describes sensory processing on an internal circle of directions
~\citep{SalahshourCouzin2026HT}. A neighbour contributes a \emph{sensory kernel},
an angular input profile centred on its bearing. For the wrapped-Gaussian kernel
used here, the width \(\sigma\) controls sensory acuity: a broader kernel gives
a less precise representation of the bearing. The agent's current heading is
encoded by a \emph{decision bump}, a localized region of elevated internal
activity on the same circle. In our square-profile baseline, \(W\) is the full
angular width of this bump's active region.

When the bump shape relaxes rapidly compared with changes in heading and the
sensory scene, HT obtains the directional interaction from the overlap of the
sensory input and decision bump~\citep{SalahshourCouzin2026HT}. Expanding both
profiles in angular Fourier series separates this interaction into perceptual
channels. Each signed coefficient \(q_n\) is the product of the corresponding
sensory coefficient \(c_n(\sigma)\) and decision coefficient \(M_n(W)\).
The radial filter \(J(r)\) sets how the interaction varies with neighbour distance,
giving the microscopic coupling \(K_n(r)=J(r)q_n\). HT derives this factorization
for both square and Gaussian decision profiles
~\citep[SM Sec.~S.2.5]{SalahshourCouzin2026HT}. For the wrapped-Gaussian sensory
kernel and signed square decision bump used as our baseline,
\begin{equation}
  \begin{gathered}
  q_n=c_n(\sigma)M_n(W),\\
  c_n=\frac{\e^{-n^2\sigma^2/2}}{\pi},\qquad
  M_n=\frac{4}{n\pi}\sin\frac{nW}{2}.
  \end{gathered}
  \label{eq:coefficients-methods}
\end{equation}
Changing the sign of \(M_n\) reverses that channel's turning response
at fixed bearing and radial coupling.

To write the resulting bearing law, consider \(N\) agents at positions
\(\rr_i\) with headings \(\phi_i\) and unit heading vector
\(\ee(\phi)=(\cos\phi,\sin\phi)\). Angles increase counterclockwise. The directed
displacement from agent \(i\) to agent \(j\) is
\(\rr_{ij}=\operatorname{MI}(\rr_j-\rr_i)=r_{ij}\ee(\theta_{ij})\), where
\(\operatorname{MI}\) is the minimum-image map on a periodic domain and the
identity on the plane.
\ HT shows that the corresponding position--heading stochastic differential equation (SDE) is
given by Eq.~\eqref{eq:particle-law}:
\begin{equation}
\begin{aligned}
  \dd\rr_i &=v_0\ee(\phi_i)\,\dd t,\\
  \dd\phi_i &=\eta\sum_{j\ne i}\sum_{n\geq1}nq_nJ(r_{ij})
  \sin[n(\theta_{ij}-\phi_i)]\,\dd t
  +\sqrt{2\Dr}\,\dd\mathcal B_i .
\end{aligned}
\label{eq:particle-law}
\end{equation}
Here \(v_0\) is speed, \(\eta\) is rotational mobility, and
\(J(r)\equiv J_R(r)\) is the signed radial interaction kernel (the subscript \(R\)
is retained in the SM when its range matters).  Further,
\(\Dr\) is rotational diffusion, and \(\mathcal B_i(t)\) are independent
standard Wiener processes.    \ Intuitively, each agent moves forward along
its current heading, turns according to the bearings and distance-dependent
influence of the neighbors it perceives, and undergoes rotational noise.  The
sensory and decision coefficients and their signs are
derived in SM Sec.~S.2 and illustrated in SM Fig.~S1.
Robustness checks with a matched von Mises sensory filter and a smoothed square decision bump appear in SM Sec.~S.7 and SM Fig.~S5.
The derivation itself uses the Fourier coefficients of a fixed, reflection-symmetric decision profile. A Gaussian decision bump can therefore replace the square profile without changing the kinetic construction. We test this replacement separately in SM Sec.~S.7.1 and collective states in SM Sec.~S.10.6.

For each HT perceptual harmonic in Eq.~\eqref{eq:particle-law}, the
microscopic coefficient \(q_n=c_nM_n\) is fixed by the sensory and decision
parameters, whereas the remaining neighbor sum depends on the agents' current
spatial arrangement.  To prepare the continuum limit, we separate these two parts
and collect the neighbor contribution of order \(n\) into the complex cue below.
This is an exact rewriting of Eq.~\eqref{eq:particle-law}, not yet an approximation. The modulus of \(C_{n,i}\) measures the strength of the \(n\)-fold bearing signal,
and its phase specifies the signal's spatial orientation.
\begin{equation}
  \begin{gathered}
  C_{n,i}=\sum_{j\ne i}J(r_{ij})\e^{\ii n\theta_{ij}},\\
  \Omega_i=\eta\sum_{n\geq1}nq_n
  \operatorname{Im}\!\left(C_{n,i}\e^{-\ii n\phi_i}\right).
  \end{gathered}
  \label{eq:complex-cue}
\end{equation}
Write a displacement as \(\bm s=s\ee(\theta_{\bm s})\).  The phase of \(C_{n,i}\)
defines an \(n\)-fold spatial orientation pattern---one directed bearing for \(n=1\),
an unoriented axis for \(n=2\), and higher-fold patterns for larger \(n\).
\ 
Eq.~\eqref{eq:complex-cue} is now in a form that can be coarse-grained. Its neighbor sum is the microscopic-particle version of a spatial convolution.
Replacing the discrete particle density by a continuous density field gives
~(SM Secs.~S.3 and S.4)
\begin{equation}
  C_n[\rho](\xx,t)=\int \dd^2\bm s\,J(s)\e^{\ii n\theta_{\bm s}}
  \rho(\xx+\bm s,t).
  \label{eq:coarse-cue-main}
\end{equation}
Its effect at a given wavelength is set by the order-\(n\)
Hankel transform of \(J\), the radial analogue of a Fourier
transform~\citep{Baddour2009}.  Thus HT's
three microscopic ingredients---sensory filtering, directional decision-making,
and distance-dependent interaction strength---remain separately visible after
coarse-graining.  This separation makes the continuum theory mechanistic. Its transport and instability coefficients can be traced to HT's individual-level
sensory and decision parameters, which are themselves linked to the underlying
neural dynamics within the HT construction~\citep{SalahshourCouzin2026HT}.  \ (SM Secs.~S.2 and~S.5.)

\begin{figure*}[p]
  \centering
  \includegraphics[width=\linewidth]{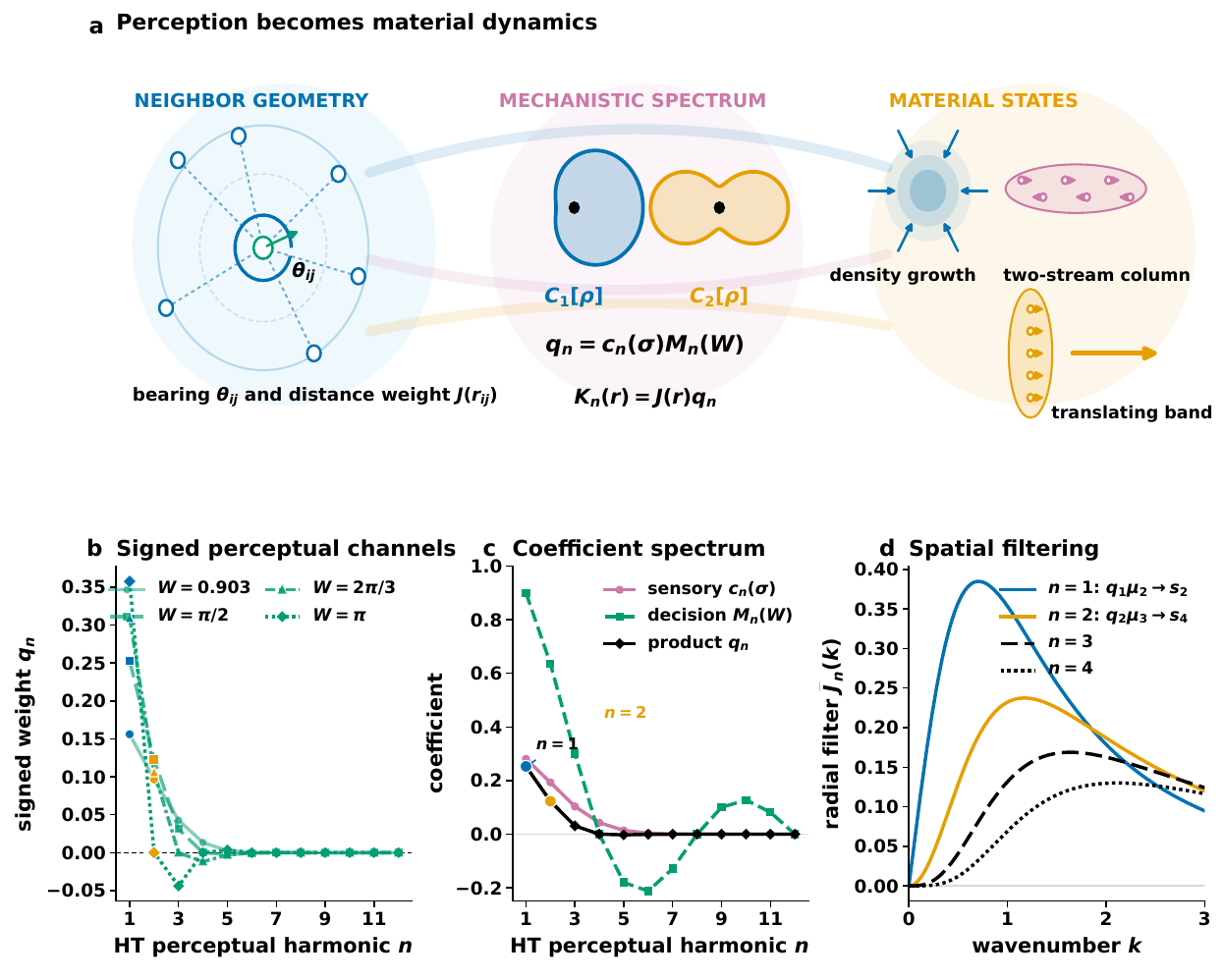}
  \caption{
  \textbf{From perception to material dynamics.}
  \textbf{(a)}, The panorama follows one construction across scales.  A focal agent
  measures neighbor bearings and distances. The resulting HT perceptual channels
  form the cue fields \(C_n[\rho]\). The continuum develops density growth, a
  bidirectional column, or a translating band without a heading-alignment torque.
  The mathematical bridge begins with the mechanistic coefficients \(q_n\)
  (Eq.~\eqref{eq:coefficients-methods})~\citep{SalahshourCouzin2026HT}, passes through the particle law and cue
  convolution in Eqs.~\eqref{eq:particle-law} and \eqref{eq:coarse-cue-main}, and
  reaches the kinetic continuum of Eqs.~\eqref{eq:kinetic-main} and
  \eqref{eq:kinetic-torque-main} and its slow density branch,
  Eq.~\eqref{eq:hydrodynamic-density}.
  \textbf{(b)}, At fixed sensory-acuity parameter \(\sigma=0.5\), the decision width
  \(W\) changes both the magnitude and sign of the perceptual coefficients \(q_n\)
  through Eq.~\eqref{eq:coefficients-methods}.
  \textbf{(c)}, Eq.~\eqref{eq:coefficients-methods} separates sensory filtering \(c_n(\sigma)\),
  directional decision-making \(M_n(W)\), and their product \(q_n\).
  This panel uses \(W=\pi/2\) and \(\sigma=0.5\).
  \textbf{(d)}, For a normalized unit-range exponential kernel, the plotted radial multiplier is
  \(\widehat J_n(k)=2\pi\mathcal J_n(k)\), where \(\mathcal J_n\) is defined in
  Eq.~\eqref{eq:hankel-methods}.  The full functions enter the nonlocal operator in
  Eq.~\eqref{eq:linear-operator-methods}. Their small-\(k\) expansions supply the
  radial moments in Eq.~\eqref{eq:longwave-coefficients-methods}.  The labels
  \(q_1\mu_2\) and \(q_2\mu_3\) denote contributions to \(s_2\) and \(s_4\), not the
  complete coefficients.
  The continuum coefficients thus remain explicitly connected to their microscopic perceptual origins.}
  \label{fig:cue}
\end{figure*}

Fig.~\ref{fig:cue}a--d previews the route across scales.  The weights \(q_n\) set the
bearing torque; the torque drives a kinetic density \(f(\xx,\phi,t)\); heading moments of \(f\) define density, polarization, and higher orientational fields; and
their slow density eigenmode gives the hydrodynamic density equation in
Eq.~\eqref{eq:hydrodynamic-density}.  The particle-to-field
derivation, heading-moment hierarchy, and Fourier--Hankel representation are given in
SM Secs.~S.3--S.5;

the complete set of radial transforms is shown in
SM Fig.~S2.

SM Sec.~S.1.1 summarizes the mathematical steps and their assumptions.

\subsection{An explicit closure turns particles into a kinetic field}
\label{sec:kinetic-closure}
\label{sec:methods-kinetic}

Passing from particles to a one-particle field requires one explicit
statistical approximation.  Let \(f^{(1)}(\xx,\phi,t)\) be the expected number of
agents per unit area per radian, and let
\(f^{(2)}(\xx,\phi;\xx+\bm s,\phi',t)\) be the corresponding density of ordered
pairs.  Applying the Fokker--Planck construction and the
Bogoliubov--Born--Green--Kirkwood--Yvon (BBGKY) reduced-density hierarchy to the particle dynamics in
Eq.~\eqref{eq:particle-law}~\citep{Risken1989,BornGreen1946,Kirkwood1946,Spohn1991}
gives the exact one-particle balance~(SM Sec.~S.3)
\begin{widetext}
\begin{equation}
\begin{aligned}
\partial_t f^{(1)}+v_0\ee(\phi)\!\cdot\!\nabla f^{(1)}
&=\Dr\partial_\phi^2f^{(1)}-\partial_\phi\mathcal G^{(2)},\\
\mathcal G^{(2)}
&=\eta\sum_{n\geq1}nq_n\!\int\!\dd^2\bm s\,\dd\phi'\,
J(s)\sin[n(\theta_{\bm s}-\phi)]
f^{(2)}(\xx,\phi;\xx+\bm s,\phi',t).
\end{aligned}
\label{eq:pair-balance-methods}
\end{equation}
\end{widetext}
This equation is exact but not closed. The evolution of the one-particle density
depends on how pairs are arranged in both position and heading.  The molecular-chaos
closure treats two agents that meet at a given instant as statistically independent
~\citep{Bertin2006,Peshkov2014},
\begin{equation}
\begin{aligned}
f^{(2)}&(\xx,\phi;\xx+\bm s,\phi',t)\\
&\simeq\frac{N-1}{N}f^{(1)}(\xx,\phi,t)
f^{(1)}(\xx+\bm s,\phi',t).
\end{aligned}
\label{eq:pair-factorization-methods}
\end{equation}
The factor \((N-1)/N\) removes self-pairs and tends to one as \(N\) grows.
A controlled regime motivating this closure has interaction range \(R_N\) and
system size \(L_N\) satisfying
\(\meanrho^{-1/2}\ll R_N\ll L_N\) and
\(\meanrho R_N^2\to\infty\): each agent samples many neighbors, while the influence
of any one neighbor becomes weak~\citep{Oelschlager1985,Sznitman1991}.  The
associated radial law is
\(J_{R_N}(r)=gR_N^{-2}j(r/R_N)\), where \(j\) is a dimensionless radial shape
normalized by \(2\pi\int_0^\infty u j(u)\,\dd u=1\), and \(g\) is its signed
integrated strength.  We do not prove convergence to molecular chaos for the
position--heading dynamics in Eq.~\eqref{eq:particle-law}. At fixed range,
Eq.~\eqref{eq:pair-factorization-methods} remains an approximation.  Its instantaneous and
finite-time predictions are compared with particle dynamics below and in
SM Sec.~S.10.3.  Substituting it into Eq.~\eqref{eq:pair-balance-methods},
integrating over the neighbor heading \(\phi'\), and dropping the asymptotically
negligible finite-\(N\) factor gives~(SM Sec.~S.3)
\begin{align}
  \partial_tf+v_0\ee(\phi)\!\cdot\!\nabla f
  &=-\partial_\phi[\Omega[f]f]+\Dr\partial_\phi^2f,
  \label{eq:kinetic-main}\\
  \Omega[f](\xx,\phi,t)
  &=\eta\sum_{n\geq1}nq_n\operatorname{Im}
  \!\left[C_n[\rho](\xx,t)\e^{-\ii n\phi}\right].
  \label{eq:kinetic-torque-main}
\end{align}
Here \(f(\xx,\phi,t)\) is number density per area per radian and
\(\rho(\xx,t)=\int_0^{2\pi}f\,\dd\phi\).  Eqs.
\eqref{eq:kinetic-main} and \eqref{eq:kinetic-torque-main} retain the full spatial

nonlocality of the perceptual cue. The turning rate at \(\xx\) depends on
  density at displaced points \(\xx+\bm s\) across the interaction range, weighted
  by \(J(s)\), rather than only on fields evaluated at \(\xx\).
SM Sec.~S.3 derives the finite-\(N\) normalization in
Eqs.~\eqref{eq:pair-balance-methods} and \eqref{eq:pair-factorization-methods}; SM Sec.~S.9.4
isolates the pair-correlation contribution discarded by the closure.

\subsection{Heading moments expose density, polarization, and orientational structure}
\label{sec:heading-moments}

The kinetic density \(f(\xx,\phi,t)\) contains an entire distribution of
headings at every point.  To turn that distribution into interpretable fields, we
Fourier transform only the heading angle~\citep{Bertin2006,Peshkov2014}:
\begin{equation}
\begin{gathered}
f_m(\xx,t)=\int_0^{2\pi}\!\e^{\ii m\phi}f(\xx,\phi,t)\,\dd\phi,\\
f(\xx,\phi,t)=\frac{1}{2\pi}\sum_{m\in\mathbb Z}f_m(\xx,t)\e^{-\ii m\phi}.
\end{gathered}
\label{eq:angular-transform-methods}
\end{equation}

In
Eq.~\eqref{eq:angular-transform-methods}, \(f_0=\rho\) is density, \(f_1\) is
polarization density, \(f_2\) is nematic density, and higher \(f_m\) resolve
successively finer heading structure.

Multiplying the kinetic equation, Eq.~\eqref{eq:kinetic-main}, by
\(\e^{\ii m\phi}\) and integrating over \(\phi\) gives
~(SM Sec.~S.4 and SM Appendix~A)
\begin{widetext}
\begin{equation}
\begin{aligned}
\partial_tf_m+\frac{v_0}{2}
(\partial_-f_{m+1}+\partial_+f_{m-1})
={}&-\Dr m^2f_m\\
&+\frac{\eta m}{2}\sum_{n\geq1}nq_n
\left(C_n[\rho]f_{m-n}-C_{-n}[\rho]f_{m+n}\right),
\end{aligned}
\label{eq:moment-hierarchy-methods}
\end{equation}
\end{widetext}
where \(\partial_\pm=\partial_x\pm\ii\partial_y\) and
\(C_{-n}=C_n^*\).  Eq.~\eqref{eq:moment-hierarchy-methods} is an exact
rewriting of the closed kinetic model. No heading-moment cutoff has yet been made.
The streaming term couples adjacent moments because a particle moving along
\(\ee(\phi)\) contains only the angular factors \(\e^{\pm\ii\phi}\). The torque term shifts \(m\) by the HT perceptual order \(n\).  SM Sec.~S.4 and
SM Appendix~A derive both projections.

Setting \(m=0\) in Eq.~\eqref{eq:moment-hierarchy-methods} gives the continuity law
~(SM Sec.~S.4)
\begin{equation}
\partial_t\rho+v_0\nabla\!\cdot\!\bm p=0,
\qquad \bm p=(\operatorname{Re}f_1,\operatorname{Im}f_1).
\label{eq:continuity-methods}
\end{equation}
Thus density cannot change by itself: it changes only when the polarization current
\(v_0\bm p\) converges or diverges.

The cue in Eq.~\eqref{eq:coarse-cue-main} is nonlocal in real space but becomes a
simple multiplier after a spatial Fourier transform
~\citep{Baddour2009} (SM Sec.~S.5 and SM Appendix~A).  Writing the spatial wavevector as
\(\bm k=k(\cos\chi,\sin\chi)\), where \(\chi\) is its direction angle,
\begin{equation}
\begin{gathered}
\widehat C_n(\bm k)=2\pi\ii^n\e^{\ii n\chi}\Jcal_n(k)\widehat\rho(\bm k),\\
\Jcal_n(k)=\int_0^\infty rJ(r)\,\mathrm J_n(kr)\,\dd r,
\end{gathered}
\label{eq:hankel-methods}
\end{equation}
where \(\mathrm J_n\) is the Bessel function of the first kind~\citep{Olver2010}.
The order-\(n\) Hankel transform \(\Jcal_n(k)\)~\citep{Baddour2009}
therefore tells us how the radial interaction law filters each
spatial wavelength in the \(n\)th HT perceptual channel.

We now perturb uniform angular disorder,
\(f_0=\meanrho\) and \(f_{m\ne0}=0\), by
\(\delta f_m=\widehat f_m\e^{\lambda t+\ii\bm k\cdot\xx}\).  Linearizing
Eq.~\eqref{eq:moment-hierarchy-methods} gives~(SM Sec.~S.5)
\begin{widetext}
\begin{equation}
\begin{aligned}
\lambda\widehat f_m&=\sum_{m'=-m_{\max}}^{m_{\max}}
\mathsf L_{mm'}\widehat f_{m'},\\
\mathsf L_{mm'}={}&-\Dr m^2\delta_{mm'}
-\frac{\ii v_0k}{2}
\left(\e^{-\ii\chi}\delta_{m',m+1}
+\e^{\ii\chi}\delta_{m',m-1}\right)\\
&+\delta_{m'0}\mathbb 1_{m\ne0}\,
\eta\pi\meanrho\,m^2q_{|m|}\ii^{|m|}\e^{\ii m\chi}
\Jcal_{|m|}(k).
\end{aligned}
\label{eq:linear-operator-methods}
\end{equation}
\end{widetext}
Each eigenvalue of Eq.~\eqref{eq:linear-operator-methods} gives a growth or decay
rate and, through its imaginary part, an oscillation frequency.  Its eigenvector
gives the accompanying mixture of density and heading moments.
The spectrum becomes independent of \(m_{\max}\) once the heading
representation is converged.  Rotation and
reflection separate a density-coupled longitudinal sector, which contains every
possible instability, from a strictly damped transverse sector for \(D_{\mathrm r}>0\).  SM Sec.~S.5 and
SM Appendix~A derive Eq.~\eqref{eq:hankel-methods}, prove
the symmetry reduction, and document cutoff convergence.

The operations leading to Eq.~\eqref{eq:linear-operator-methods}---a
one-particle closure, angular projection, and spatial transform---are standard
kinetic-theory tools~\citep{Risken1989,Spohn1991,Bertin2006,Peshkov2014}.
The methodological result is the operator they produce for
HT. Every matrix element retains the indexed perceptual coefficient \(q_n\) and the
corresponding radial transform \(\widehat J_n(k)\).  Local vision-cone
hydrodynamics~\citep{Peruani2017} does not retain this complete finite-wavenumber
spectrum, while the nonlocal visual theory of Ref.~\citep{ItoUchida2025} uses a
heading-averaging microscopic update.  Preserving the HT structure is what permits
the channel-specific and radial-filtering predictions below without refitting a
macroscopic transport coefficient.

\section{Perceptual feedback creates hydrodynamic instabilities}

\subsection{Density--bearing feedback destabilizes uniform disorder}
\label{sec:density-instability}
\label{sec:methods-onset}

\begin{figure}[!t]
  \centering
  \includegraphics[width=\linewidth]{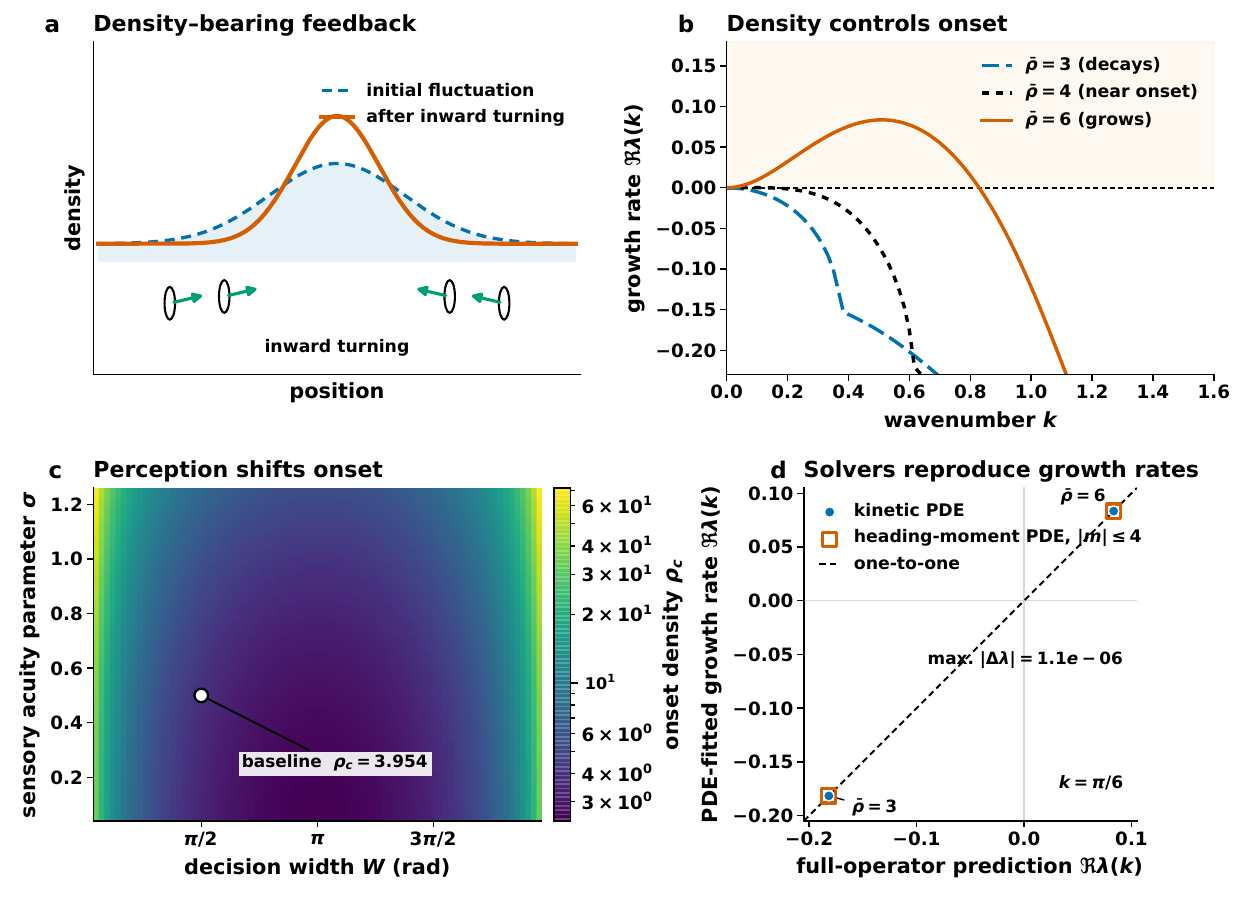}
  \caption{
  \textbf{Density--bearing feedback sets the onset of organization.}
  \textbf{(a)}, A density excess generates the coarse-grained \(n=1\) bearing cue
  \(C_1[\rho]\) through Eq.~\eqref{eq:coarse-cue-main}.  The resulting torque
  turns agents inward (Eq.~\eqref{eq:kinetic-torque-main}), and self-propulsion
  carries them into the excess. This feedback enters density evolution through
  Eq.~\eqref{eq:hydrodynamic-density}.
  \textbf{(b)}, The leading eigenvalue
  \(\lambda(k)\) of Eq.~\eqref{eq:linear-operator-methods} gives the growth or decay rate
  \(\Re\lambda(k)\); positive values grow.  Raising \(\meanrho\) from 3 to 6 crosses
  the onset.
  \textbf{(c)}, Eq.~\eqref{eq:threshold} maps the onset
  density (color) across decision width \(W\) (horizontal axis) and
  sensory-acuity parameter \(\sigma\) (vertical axis).  The
  marked baseline is \(W=\pi/2\), \(\sigma=0.5\), and \(\rho_c=3.954\).
  Larger \(\sigma\) means lower sensory acuity.
  \textbf{(d)}, 
  PDE-fitted growth rates (vertical axis) are plotted against full-operator
  predictions (horizontal axis) from Eq.~\eqref{eq:linear-operator-methods}.
  Filled circles use the kinetic PDE, Eqs.~\eqref{eq:kinetic-main} and
  \eqref{eq:kinetic-torque-main}; open squares use the truncated heading-moment PDE
  (Eq.~\eqref{eq:moment-hierarchy-methods}, \(\lvert m\rvert\leq4\)).  Both follow
  the one-to-one line.  Panels b and
  d use \(v_0=\eta=1\), \(W=\pi/2\), \(\sigma=0.5\), \(\Dr=0.25\), and a normalized
  range-one exponential kernel.  
  Numerical details are given in SM Sec.~S.10.2.}
  \label{fig:onset}
\end{figure}

On the infinite plane, or when a radial kernel is supported inside the inscribed
disk of a periodic box, uniform density is rotationally isotropic and every nonzero
bearing cue vanishes~(Eq.~\eqref{eq:coarse-cue-main}, \(n\ne0\)).
Positive rotational diffusion then damps all spatially
uniform polar, nematic, and higher
heading moments \(f_m\)~(Eq.~\eqref{eq:moment-hierarchy-methods}, at
\(\nabla f_m=0\) and \(C_n=0\)).  The first possible instability must
therefore couple density to orientation. A density excess creates a bearing cue,
the cue turns agents inward, and self-propulsion increases the excess
(Fig.~\ref{fig:onset}a).
(SM Secs.~S.5 and~S.6.)

Hydrodynamics describes wavelengths much larger than the interaction
range~\citep{CrossHohenberg1993,Marchetti2013}.  At those wavelengths density is slow because Eq.~\eqref{eq:continuity-methods}
conserves it, whereas every non-density heading moment in
Eq.~\eqref{eq:moment-hierarchy-methods} retains the finite damping rate
\(\Dr m^2\).  Expanding the Hankel multipliers in Eq.~\eqref{eq:hankel-methods} at
small \(k\) and eliminating these faster moments gives the slow density eigenvalue
~(SM Sec.~S.6 and SM Appendix~A)
\begin{align}
\lambda_0(k)&=s_2k^2+s_4k^4+\order(k^6),\nonumber\\
s_2&=\frac{v_0}{\Dr}
\left(\frac{\eta\pi\meanrho q_1\mu_2}{2}-\frac{v_0}{2}\right),\nonumber\\
s_4&=-\frac{v_0\eta\pi\meanrho q_1\mu_4}{16\Dr}
+\frac{v_0^2\eta\pi\meanrho q_2\mu_3}{16\Dr^2}
-\frac{v_0^2s_2}{16\Dr^2}-\frac{s_2^2}{\Dr},
\label{eq:longwave-coefficients-methods}
\end{align}
where \(\mu_p=\int_0^\infty r^pJ(r)\,\dd r\).  Inverse spatial Fourier
transformation, with \(D_{\mathrm{eff}}=-s_2\), gives~(SM Sec.~S.6)
\begin{equation}
  \begin{gathered}
  \partial_t\delta\rho
  =D_{\mathrm{eff}}\nabla^2\delta\rho
  +s_4\nabla^4\delta\rho+\order(\nabla^6\delta\rho),\\
  D_{\mathrm{eff}}=\frac{v_0}{2\Dr}
  \left(v_0-\eta\pi\meanrho q_1\mu_2\right),
  \end{gathered}
  \label{eq:hydrodynamic-density}
\end{equation}
where \(\delta\rho=\rho-\meanrho\).

SM Sec.~S.6 derives
Eq.~\eqref{eq:longwave-coefficients-methods}, shows that it recovers the noninteracting limit, and states
the conditions under which the gradient expansion is controlled.
A negative \(D_{\mathrm{eff}}\) amplifies sufficiently
long-wavelength density fluctuations.  The onset condition is therefore
\begin{equation}
  \boxed{\eta\pi\meanrho q_1\mu_2>v_0,\qquad
  \rho_c=\frac{v_0}{\eta\pi q_1\mu_2}},
  \label{eq:threshold}
\end{equation}
for \(\Dr>0\) and \(q_1\mu_2>0\).    Above threshold, a slightly denser region generates
the coarse-grained \(n=1\) bearing cue \(C_1[\rho]\), the continuum
counterpart of HT's first perceptual harmonic.  This cue turns more agents inward
and further increases the local density (Fig.~\ref{fig:onset}a,b).  This
  
  bearing-mediated density feedback is a constant-speed, torque-driven
  instability.  Its negative-diffusivity structure has precedents in alignment-free
  vision-cone hydrodynamics~\citep{Peruani2017} and in torque-driven active phase
  separation~\citep{ZhangAlertYanWingreenGranick2021}.  It remains distinct from
  conventional motility-induced phase separation, in which persistent collisions
  cause self-trapping and an effective reduction of motility with density
  ~\citep{FilyMarchetti2012,RednerHaganBaskaran2013,CatesTailleur2015}.  Eq.~\eqref{eq:threshold} connects this instability to HT's perceptual parameters:
  \(q_1=c_1(\sigma)M_1(W)\) keeps sensory acuity and decision geometry
  separately visible, while \(\mu_2\) carries the signed radial interaction law.
  The equation thereby identifies the macroscopic onset with the same first HT
  perceptual channel that drives direct approach at the individual level
  ~\citep{SalahshourCouzin2026HT}.

(SM Sec.~S.6.)

For \(\Dr>0\), spreading and concentration in
Eq.~\eqref{eq:hydrodynamic-density} share \(1/\Dr\), so it cancels from their balance
and leaves \(\rho_c\) unchanged.  At \(\Dr=0\), heading moments do not relax, so the
reduction to Eq.~\eqref{eq:hydrodynamic-density} fails and stability must be read
from the full operator.  The
Fourier--Hankel conventions, finite-wavenumber matrix, and proof that any instability
belongs to the longitudinal density-coupled sector are given in SM Sec.~S.5.

Eq.~\eqref{eq:threshold} maps sensory acuity, decision width, speed,
mobility, and interaction range onto an onset density; Fig.~\ref{fig:onset}c shows
that map over \((W,\sigma)\).  Direct kinetic integration and the
heading-moment hierarchy reproduce the full-operator growth and decay rates in
Fig.~\ref{fig:onset}d.  The first nonlinear correction to this
near-threshold theory is developed below, after the finite-wavelength structure has
been established.

\begin{figure}[!t]
  \centering
  \includegraphics[width=\linewidth]{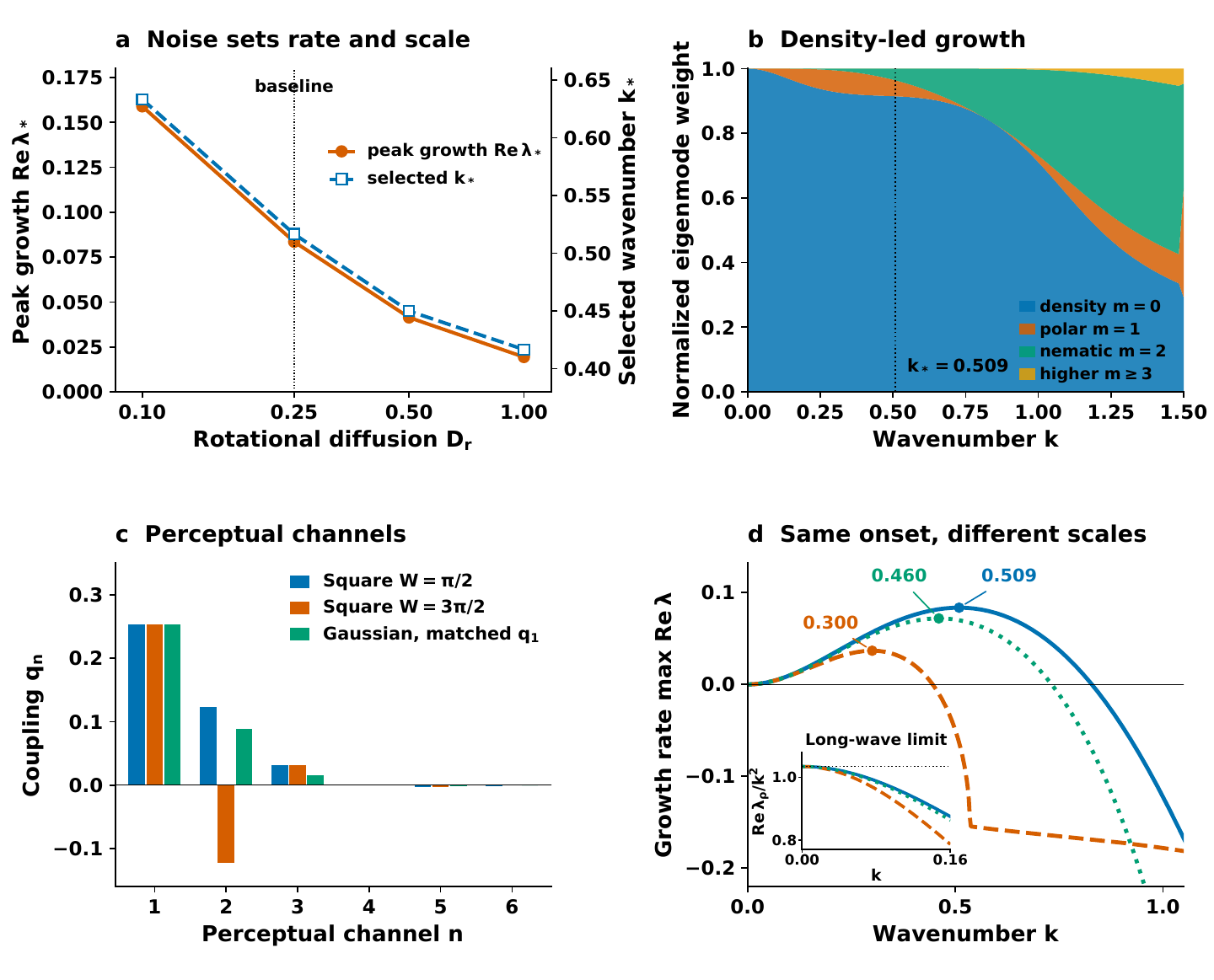}
  \caption{
  \textbf{The full nonlocal operator selects the initial pattern scale.}
  \textbf{(a)}, At \(\meanrho=6\), rotational diffusion changes both the peak growth
  rate and the selected wavenumber predicted by
  Eq.~\eqref{eq:linear-operator-methods}, even though the long-wave threshold
  (Eq.~\eqref{eq:threshold}) is independent of \(\Dr\) for \(\Dr>0\).
  \textbf{(b)}, At each \(k\), the leading eigenvector of Eq.~\eqref{eq:linear-operator-methods}
  is resolved in the heading-moment basis defined by
  Eq.~\eqref{eq:angular-transform-methods}.  The plotted weights are the squared
  magnitudes of its components, with the \(\pm m\) pair combined for \(m\geq1\)
  and all weights normalized to sum to one.  Near the baseline maximum the
  linear eigenmode is density-led, with smaller polar and nematic components
  coupled to density through Eq.~\eqref{eq:moment-hierarchy-methods}.
  \textbf{(c)}, Complementary square decision widths preserve the odd perceptual
coefficients and reverse the even ones (Eq.~\eqref{eq:coefficients-methods}).
The Gaussian decision profile matches the first coefficient and full
width at half maximum of the \(W=\pi/2\) square profile, with a different
higher-channel spectrum.
\textbf{(d)}, The maximum real eigenvalue of Eq.~\eqref{eq:linear-operator-methods}
for the three spectra in (c). Points label the refined fastest wavenumbers.
Inset, the density-branch growth divided by \(k^2\) approaches the shared
\(s_2=1.035\) (Eq.~\eqref{eq:longwave-coefficients-methods}), while the curves
separate at finite \(k\). All three profiles have \(\rho_c=3.954\).
The Gaussian selects \(k_*=0.460\), compared with \(0.509\) for the
width-matched square profile, giving a larger initial spacing.
Panels a and b use the baseline parameters of Fig.~\ref{fig:onset}; the vertical
line in (b) marks the same refined baseline maximum as in (d). Panels c and d use
\(\meanrho=6\), \(D_{\rm r}=0.25\), \(\sigma=0.5\), \(v_0=\eta=1\), and
\(J(r)=e^{-r}/(2\pi)\), with 16 perceptual channels and heading moments
through \(|m|=28\). The wavelengths concern linear growth, before nonlinear
pattern evolution (SM Secs.~S.6 and S.7.1).}
  \label{fig:finite-k}
\end{figure}

\subsection{Finite wavelengths select what grows}

Eq.~\eqref{eq:threshold}
tells us only whether very long density waves can grow.  To determine what appears
first, we evaluate the full linearized heading-moment operator in
Eq.~\eqref{eq:linear-operator-methods}, at each spatial wavenumber \(k\).  Its
leading eigenvalue gives the growth rate, while its eigenvector gives the mixture
of density and orientational fields in that growing disturbance.  The value of
\(k\) that maximizes \(\operatorname{Re}\lambda(k)\) is the fastest-growing
wavenumber; its inverse sets the initial pattern spacing.  Although \(\Dr\) cancels from \(\rho_c\),
it changes this finite-wavelength calculation.  At \(\meanrho=6\), raising \(\Dr\) from
0.1 to 1 lowers the sampled peak growth rate from 0.159 to 0.0194 and shifts the
fastest wavenumber from 0.633 to 0.417 (Fig.~\ref{fig:finite-k}a and
SM Fig.~S3b).  Noise therefore changes the pace and length scale of organization even
when it leaves the long-wave boundary unchanged. Here stronger noise produces a slower, more widely spaced pattern.  (SM Secs.~S.5 and S.6.)

At each \(k\), the corresponding eigenvector decomposes
the growing disturbance among the heading moments.  We square the component
magnitudes and normalize them to sum to one. These weights describe the composition
of the linear disturbance.  At the baseline
 fastest-growing wavenumber,  the eigenvector's normalized weights are
91.4\% density, 4.9\% longitudinal polarization, and 3.7\% nematic structure;
higher heading moments carry negligible weight at that wavelength
(Fig.~\ref{fig:finite-k}b and SM Sec.~S.5).
Density is therefore the
largest component, but it cannot grow alone: in
Eq.~\eqref{eq:continuity-methods}, density changes through the divergence of the
polarization current.  The 4.9\% longitudinal-polarization component is thus
dynamically essential, while the nematic component records the accompanying
twofold orientational structure.
(SM Sec.~S.5.)

\subsection{Higher perceptual channels enter at progressively higher derivative order}
\label{sec:higher-harmonic-entry}

We next ask how HT's microscopic perceptual channels shape these spatial
scales.  Here \(n\) labels an HT perceptual harmonic, \(k\) labels a spatial
wavenumber, and \(m\) labels a heading moment.  The
first HT perceptual harmonic changes density
transport at order \(k^2\).  The second HT perceptual harmonic first contributes
explicitly at order \(k^4\).  For a spatially isotropic convolution, \(\Dr>0\), an
isolated small-\(k\) density branch, and sufficient absolute radial moments,
returning an order-\(n\) cue to density requires \(n\) successive couplings through the streaming term in Eq.~\eqref{eq:moment-hierarchy-methods} (each coupling changes the heading-moment index by one and contributes one power of \(k\)).
The cue supplies \(n\) powers of \(k\), and those streaming steps supply
another \(n\): the round trip from density through the order-\(n\) response and
back to density therefore begins with \(k^{2n}\).
Consequently, the term linear in an infinitesimally activated \(q_n\) generically
first appears at order \(k^{2n}\), provided its leading radial moment
\(\mu_{n+1}\) is nonzero.  A cancellation delays the entry to a higher even power.
This rule identifies the first possible order; it does not determine the sign of
that contribution.  The explicit leading term is
\begin{equation}
\Delta_n\lambda_0(k)=
\frac{\eta\pi\meanrho q_n\mu_{n+1}n^2v_0^n}
{2^{2n-1}\Dr^n(n!)^3}k^{2n}+\order(k^{2n+2}).
\label{eq:harmonic-entry-methods}
\end{equation}
Eq.~\eqref{eq:harmonic-entry-methods} makes the microscopic connection
transparent. The sign and strength come from the HT coefficient \(q_n\) and the
radial moment \(\mu_{n+1}\), whereas the power \(2n\) follows from geometry and
streaming.  HT therefore supplies the indexed coefficient \(q_n\), including its
sign and sensory interpretation. The continuum calculation supplies the exponent
\(2n\).  This gives a conditional channel-to-leading-derivative rule.  A local
theory can recover the same order-by-order attribution only if it is supplied with
the corresponding order-resolved HT inputs.  SM Sec.~S.6 gives the coefficient of this term, the
complete fourth-order coefficient \(s_4\), its
reduction to the noninteracting limit, and the conditions
under which fourth order regularizes the instability.  If those conditions fail,
the full nonlocal operator, not the truncated gradient equation in
Eq.~\eqref{eq:hydrodynamic-density}, must select the scale.

Thus the first HT perceptual harmonic controls long-wave onset, whereas
higher channels first appear at progressively higher spatial-derivative order.
Their actual effect on a selected wavelength must still be read from the full
nonlocal operator.
(SM Sec.~S.6 and SM Appendix~A.)

A direct comparison makes this distinction visible. For the square decision
profile in Eq.~\eqref{eq:coefficients-methods}, complementary widths \(W\) and
\(2\pi-W\) preserve every odd perceptual channel and reverse every even one
(Fig.~\ref{fig:finite-k}c). The two populations therefore share the same first
channel and long-wave threshold, while their shorter-scale response differs.
At the baseline density, \(W=\pi/2\) and \(3\pi/2\) give fastest wavenumbers
\(k_*=0.509\) and \(0.300\), respectively, in the full operator
(Fig.~\ref{fig:finite-k}d). Their initial pattern spacings are thus 12.3 and
20.9 interaction lengths. Two populations can become unstable at the same
long-wave threshold yet favor different initial spatial scales because they
perceive differently (SM Sec.~S.6).

Two results show why
the initial pattern scale must be read from the converged full operator rather than
inferred from Eq.~\eqref{eq:hydrodynamic-density}.  First, a low heading-moment cutoff can
produce a prominent oscillatory leading branch, but it disappears after convergence;
the fastest-growing converged linear eigenmode in the demanding low-noise case is stationary
(SM Fig.~S4).  Second, a kernel with short
attraction and longer-range repulsion is stable as \(k\to0\) yet becomes unstable at
finite wavelength at density 24, with sampled maximum \(k=3.1\) and growth rate
0.721 (SM Fig.~S6).  Cutoff convergence removes a spurious oscillatory
interpretation. The finite-wavelength result shows physically that long-wave stability does not
exclude a growing finite-spacing pattern.  (SM Secs.~S.5, S.7, and~S.10.1.)

Finite-wavelength instabilities are not unique to this model; generic
nonlocal torque field theories can also produce long- and short-wave growth
~\citep{SeseSansaLevisPagonabarraga2022}.  Here the sign, growth band, and eigenvector composition follow from HT's
signed coefficients together with the Hankel transforms of the radial law.  A local
long-wave closure cannot make that attribution outside its controlled small-\(k\)
window.

Matched smooth sensory and decision profiles move the baseline threshold by only
about one percent, whereas radial-kernel shape and interaction range can change both
the threshold and fastest wavelength.  SM Video~1 illustrates how the matched
profiles affect early pattern spacing and growth rates; SM Video~2
shows how radial filtering changes the visible pattern scale.  SM Sec.~S.7 and
SM Figs.~S5--S6 give the corresponding profile, kernel,
range, and independent-transform comparisons. SM Figs.~S2--S4 give the full
transform, dispersion, and convergence spectra.  In physical terms, the
onset mechanism is robust to smoothing of the sensory profiles, whereas the radial
interaction law strongly controls the spacing of the emerging pattern.

A Gaussian decision bump gives a further test, motivated by the smooth fly compass-neuron activity profiles examined in harmonic spectroscopy~\citep{Salahshour2026Spectroscopy}. Matching its full width at half maximum to the square width and its gain to \(q_1\) preserves \(\rho_c=3.954\), but increases the fastest-growing spacing from 12.34 to 13.65 interaction lengths (Fig.~\ref{fig:finite-k}c,d). If both \(q_1\) and \(q_2\) are matched, the dispersion curves nearly coincide instead. The comparison shows how the retained perceptual channels determine the collective response across different profile shapes (SM Sec.~S.7.1 and Fig.~S7a--c).

A neighbor's finite apparent size introduces a second, geometric filter. For an unoccluded, reflection-symmetric image, its angular Fourier coefficient \(S_n(r)\) multiplies the radial kernel in each cue, \(J(r)\mapsto J(r)S_n(r)\)~\citep{Salahshour2026Spectroscopy}. The same nonlocal operator and conditional \(k^{2n}\) rule follow with channel-specific radial moments. In a controlled comparison with the same radial support and total cue weight, finite extent raises the onset density by \(0.87\%\) and the fastest-growing spacing by \(2.45\%\). These are modest but resolved geometric effects, without changing sensory acuity or the decision profile (SM Sec.~S.7.2 and Fig.~S7d--f).

\subsection{The first nonlinear correction describes early mode coupling}
\label{sec:weakly-nonlinear}

The linear operator determines which infinitesimal disturbance grows,
but it cannot describe how growing spatial modes begin to interact.  Close to the
threshold, write \(\meanrho=\rho_c(1+\epsilon)\) and
\(u=\rho-\meanrho\).  Polarization still relaxes faster than density, so expanding
the first cue, solving the \(m=1\) row of
Eq.~\eqref{eq:moment-hierarchy-methods} quasistatically, and substituting the result
into the continuity law, Eq.~\eqref{eq:continuity-methods}, gives
~(SM Sec.~S.8)
\begin{equation}
\begin{gathered}
\begin{aligned}
\partial_tu={}&\nabla^2\!\left[-D_a\epsilon u-\frac{\zeta}{2}u^2
+s_4^c\nabla^2u\right]\\
&+\text{higher-order terms},
\end{aligned}\\
D_a=\frac{v_0^2}{2\Dr},\quad
\zeta=\frac{v_0\eta\pi q_1\mu_2}{2\Dr},
\end{gathered}
\label{eq:weak-nonlinear-methods}
\end{equation}
where \(s_4^c\) is Eq.~\eqref{eq:longwave-coefficients-methods} evaluated at
threshold.  The quadratic term transfers density from an initially excited
spatial Fourier mode to its second spatial Fourier mode while conserving total particle
number.  Eq.~\eqref{eq:weak-nonlinear-methods} describes only early,
long-wavelength growth near onset.  It does not predict coexistence, transition
order, saturation, or the mature columns considered later.  SM Secs.~S.8 and~S.10.2
derive and compare this reduction.

\section{\texorpdfstring{Consistency across scales}{Consistency across scales}}

\subsection{\texorpdfstring{Comparisons between levels of description}{Comparisons between levels of description}}

\begin{figure*}[p]
  \centering
  
  \includegraphics[width=0.99\linewidth]{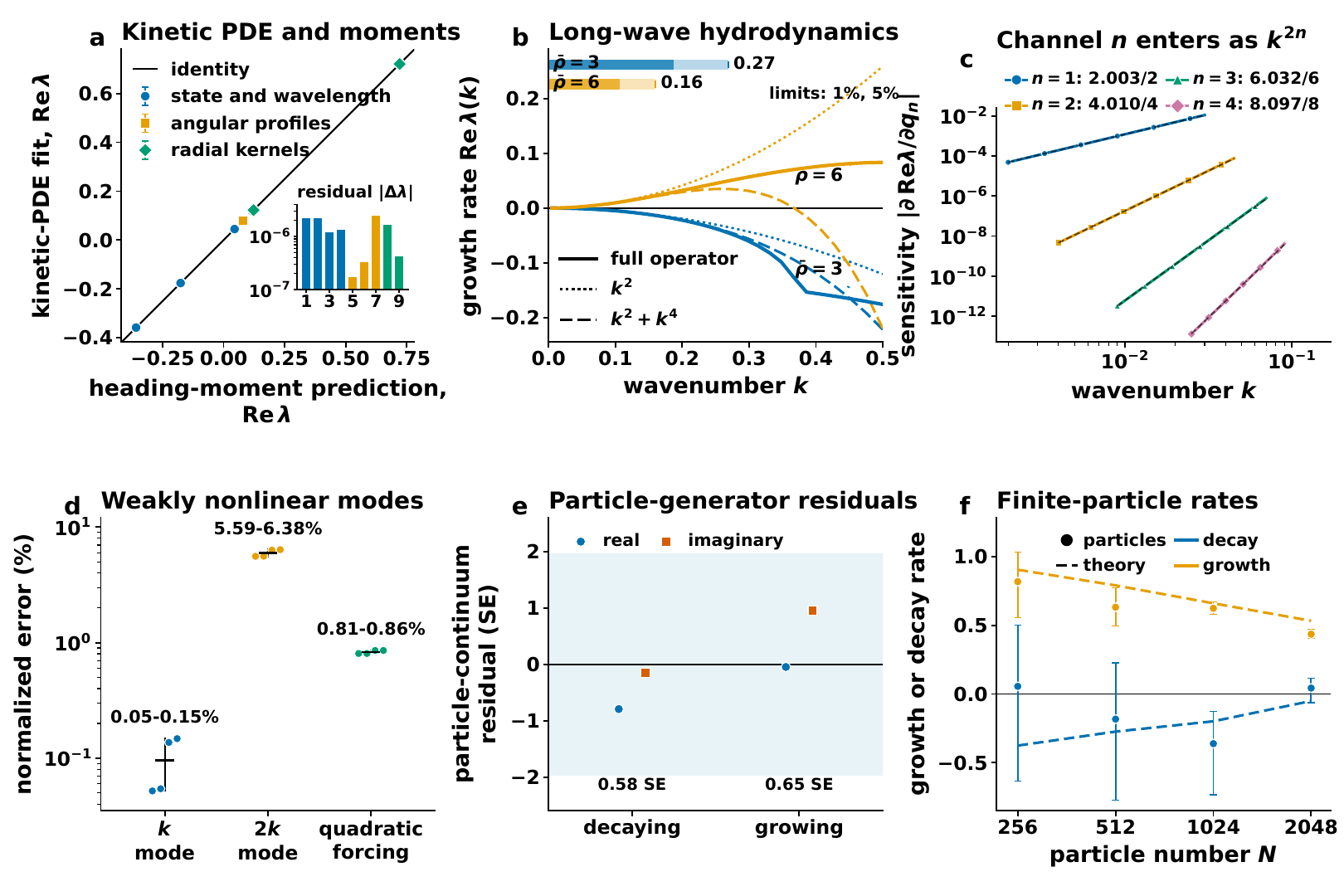}
  
  \caption{
  \textbf{Independent comparisons test each reduction at the scale
  where it is used.}
  \textbf{(a)}, Nine growth and decay rates measured from the kinetic PDE
  (Eqs.~\eqref{eq:kinetic-main} and
  \eqref{eq:kinetic-torque-main}) agree with the
  heading-moment operator in Eq.~\eqref{eq:linear-operator-methods}. The inset resolves the remaining absolute differences.
  \textbf{(b)}, The full density branch of
  Eq.~\eqref{eq:linear-operator-methods} is compared with the \(k^2\) and
  \(k^2+k^4\) forms obtained from Eqs.~\eqref{eq:longwave-coefficients-methods}
  and \eqref{eq:hydrodynamic-density}.  Dark and pale bars give the contiguous 1\%
  and 5\% validity intervals, measured contiguously from the smallest
  sampled \(k\), showing that the gradient expansion is controlled only at long
  wavelengths.
  \textbf{(c)}, At \(q_n=0\), the absolute left--right eigenvalue sensitivities
  \(\lvert\partial\Re\lambda/\partial q_n\rvert\) recover the \(k^{2n}\) exponents
  in Eq.~\eqref{eq:harmonic-entry-methods}; the dashed guides test exponents, not
  prefactors.
  \textbf{(d)}, Across four near-threshold runs, the full kinetic PDE is compared with
  Eq.~\eqref{eq:weak-nonlinear-methods} for the fundamental and generated second
  spatial modes; independently isolated kinetic forcing is compared with the
  predicted gap-rescaled quadratic coefficient.  The normalized errors are
  0.05--0.15\%, 5.59--6.38\%, and 0.81--0.86\%, respectively. Dots show runs and bars show medians.
  \textbf{(e)}, For growing and decaying one-particle states, the infinitesimal generator
  induced by Eq.~\eqref{eq:particle-law}, given explicitly in SM Eq.~(92), is
  projected onto the tested spatial and heading mode and compared with the drift
  from Eq.~\eqref{eq:linear-operator-methods}.  Real and imaginary
  particle-minus-continuum residuals are shown in standard-error units. The shaded band is \(\lvert z\rvert\leq1.96\).
  \textbf{(f)}, Along the many-neighbor Kac sequence, coherent particle rates with 95\%
  bootstrap intervals~\citep{Efron1979} are compared with the finite-\(N\) predictions defined by
  the scaling and projection in SM Eqs.~(89) and (91).  Growth is resolved at
  all four sizes. Among the decay tests, only \(N=1{,}024\) is resolved, while the
  intervals at \(N=256,512,\) and \(2{,}048\) cross zero.  This tests sign recovery,
  not quantitative asymptotic convergence.  Panels a--d test
  the closed continuum and its reductions; panels e--f examine
  the particle-to-continuum comparison under the
  prescribed one-particle state and Kac scaling.  Protocols and algebraic checks
  are given in SM Sec.~S.10.}
  \label{fig:validation}
\end{figure*}

The derivation
connects the HT particle law, the closed kinetic continuum, its heading-moment
representation, and the local long-wave density equations.  Because agreement
within one representation does not establish the next scale, each connection is
examined separately (Fig.~\ref{fig:validation}a--f).

\nocite{Sobol1967,Owen1997}

\subsubsection{\texorpdfstring{Algebraic consistency of the projected operator}{Algebraic consistency of the projected operator}}

Direct angular quadratures reproduce the streaming, torque, and Fourier--Bessel
projections in Eqs.~\eqref{eq:moment-hierarchy-methods} and \eqref{eq:hankel-methods}.
Rotating the wavevector and separating reflection sectors preserve the matrix
spectra and longitudinal reduction to numerical precision. These comparisons
check the algebra independently of time integration.
\ (SM Appendix~A and
SM Sec.~S.10.)

\subsubsection{\texorpdfstring{Agreement between the heading-moment operator and the kinetic equation}{Agreement between the heading-moment operator and the kinetic equation}}

The kinetic partial differential equation (PDE) in Eqs.~\eqref{eq:kinetic-main} and
\eqref{eq:kinetic-torque-main} evolves the complete distribution over position and
heading.  The heading-moment operator in
Eq.~\eqref{eq:linear-operator-methods} represents the same closed continuum model
after a Fourier expansion in heading.  The two calculations use independent
angular discretizations and radial transforms.  Their agreement is assessed in
nine growing and decaying cases spanning density, wavelength, smooth sensory and
decision profiles, a Gaussian kernel, and the mixed-kernel finite-wavelength
instability.  All nine kinetic-PDE rates lie on the operator prediction in
Fig.~\ref{fig:validation}a. {}  Spatial, heading, and timestep refinement preserve the
agreement.  Thus the two representations of the closed continuum dynamics are
equivalent within numerical accuracy.  The particle-to-continuum closure is tested separately in the particle comparisons below.

(SM Secs.~S.10.1--S.10.2 and SM Fig.~S12.)

\subsubsection{\texorpdfstring{Range of validity of the local density equation}{Range of validity of the local density equation}}

The local density equation in Eq.~\eqref{eq:hydrodynamic-density} is a
small-\(k\) expansion of the full nonlocal continuum operator.  Fig.~
\ref{fig:validation}b measures where that approximation is accurate.  At \(\meanrho=6\), retaining the
\(k^2+k^4\) terms gives the full density-branch growth rate to within 5\% through \(k=0.160\).  The approximation fails near
the visibly fastest-growing wavenumber, \(k\simeq0.5\).  The local equation
therefore explains long-wave onset, whereas the full nonlocal operator must predict
the visible spacing away from threshold.  

Infinitesimal activation
of each HT perceptual coefficient \(q_n\) gives the corresponding change in the
density eigenvalue.  The fitted powers for
\(n=1,\ldots,4\) are \(k^2,k^4,k^6\), and \(k^8\), respectively
(Fig.~\ref{fig:validation}c).  The \(k^{2n}\) relation is therefore present in the full
operator and is not introduced by truncating the density equation.  

(SM Secs.~S.6 and~S.10.2.)

\subsubsection{\texorpdfstring{Early nonlinear mode coupling}{Early nonlinear mode coupling}}

Beyond the initial exponential regime, we compare the weakly nonlinear density equation in
Eq.~\eqref{eq:weak-nonlinear-methods} with the full kinetic PDE in four
near-threshold runs.  The reduced equation reproduces the fundamental spatial Fourier mode with less than
0.2\% normalized error and the generated second mode with about 6\% error
(Fig.~\ref{fig:validation}d). An independent opposite-amplitude calculation also
recovers the sign and strength of the quadratic forcing.  This agreement
covers the early interaction of long spatial modes near threshold. Mature columns and strongly nonlinear motion lie outside this approximation.

(SM Secs.~S.8 and~S.10.2.)

\subsubsection{\texorpdfstring{Instantaneous particle-to-continuum correspondence}{Instantaneous particle-to-continuum correspondence}}

The particle-to-continuum step is examined
directly by evaluating their instantaneous average evolution for a
configuration drawn from the one-particle state assumed by the kinetic closure.
Independent configurations are sampled in one growing and one decaying state
~\citep{Sobol1967,Owen1997}, before any time step is taken. The particle generator
includes self-propulsion, the pairwise HT-derived torque, rotational drift,
and finite-\(N\) normalization. Projecting its heading-moment components onto
the conserved density branch resolves no particle--continuum discrepancy in
either state (Fig.~\ref{fig:validation}e).  The observed agreement supports the
instantaneous correspondence for these configurations.  Correlations that develop
later and the upstream neural derivation of \(q_n\) lie outside this comparison.
{}
\ (SM Sec.~S.10.3.)

\subsubsection{\texorpdfstring{Finite-time particle dynamics and sampling noise}{Finite-time particle dynamics and sampling noise}}
\label{sec:methods-kac}

Instantaneous generator agreement does not establish finite-time particle
dynamics.  In the fixed-range calculation, the imposed spatial Fourier
density signal is at or below the particle shot-noise floor for the two smallest
systems.  Averaging its positive magnitude can then make decay toward the noise
floor appear to be growth.  Correcting this estimator bias removes the apparent
evidence for growth, but the weak signal cannot distinguish remaining
finite-coordination or pair-correlation effects.  The signal is too weak to assess the closure quantitatively.

(SM Sec.~S.10.3, Table~S13 and SM Fig.~S14.)

A many-neighbor, weak-pair Kac sequence provides a
finite-time comparison at increasing \(N\).  As \(N\) increases, each particle samples more neighbors
while the strength of each pair decreases.  The particles are initialized in the
full predicted right eigenmode, and the matching left eigenvector isolates the
coherent response from particle noise.  The growing sign is resolved at all four
sizes in Fig.~\ref{fig:validation}f.  Decay is also resolved at \(N=1{,}024\), whereas the predicted near-neutral decay at
\(N=2{,}048\) remains unresolved.  The sequence therefore supports the predicted
growth sign and one resolved decay, but the remaining offsets and broad intervals
do not establish quantitative convergence to the continuum limit.
(SM Sec.~S.10.3 and Table~S8.)  SM Fig.~S13 shows separately how the finite periodic box changes the allowed spatial
wavelengths and can prevent an otherwise growing disturbance from fitting in the domain.

Numerical procedures are given in SM Sec.~S.10.

\section{Collective motion without heading alignment}
\label{sec:collective-motion}
\label{sec:methods-ordered}
A growing density pattern need not carry a net current.  We first examine
whether the one-particle continuum preserves HT's bidirectional columns
(Sec.~\ref{sec:attractive-column}), then test a prescribed two-stream response for selecting one direction (Sec.~\ref{sec:pair-selection}).  HT also predicts a distinctive collective state under pure repulsion: individuals move transverse to the band's long axis~\citep{SalahshourCouzin2026HT}. We test this prediction in the one-particle continuum, examining the collective current and the geometry of its density patterns (Sec.~\ref{sec:results-pure-repulsion}).  We use the full kinetic equations, Eqs.~\eqref{eq:kinetic-main} and \eqref{eq:kinetic-torque-main}, because these nonlinear states lie beyond the long-wave density approximation in Eq.~\eqref{eq:hydrodynamic-density} (SM Secs.~S.9 and S.10).

\subsection{Beyond onset: the one-particle continuum sustains a bidirectional moving column}
\label{sec:attractive-column}

HT's even perceptual harmonics can organize a common axis while treating its
two directions equivalently, consistent with the bidirectional columns reported
for constant, infinite-range attraction~\citep{SalahshourCouzin2026HT}.
HT also reports unidirectional columns when attraction decays with distance or
includes a short repulsive core that induces collision avoidance
~\citep{SalahshourCouzin2026HT}.  These radial laws permit spontaneous direction
selection without imposing a preferred direction externally (SM Sec.~S.9.4).  
The product-closed continuum determines which parts of that
phenomenology survive the one-particle closure.

All continuum calculations in this
section integrate the nonlocal kinetic continuum in
Eqs.~\eqref{eq:kinetic-main} and \eqref{eq:kinetic-torque-main}
from analytical continuum profiles; additional comparisons use smooth initial fields reconstructed from an HT particle configuration.

(SM Secs.~S.9 and~S.10.4.)

Continuum bands without microscopic heading alignment are already known,
including nematic bands and moving locally polar structures in a vision-cone
theory~\citep{Peruani2017}.  The question here is more specific: when the retained
HT perceptual spectrum is coarse-grained under product closure, does an initialized
common axis persist, and does either counterpropagating stream grow at the expense
of the other?

The HT parameter values are \(v_0=2\), \(\eta=0.1\), \(D_{\mathrm r}=0\),
\(W=0.9033\), \(\sigma=0.1\)~\citep{SalahshourCouzin2026HT}.
For computation,
lengths and times are divided by 62.5 and 31.25, respectively, retaining the
small physical mobility in the rescaled coupling. All trajectory times below
are reported in HT's physical units.  We initialize a
nematic column with two antipodal heading peaks and perturb their populations by
1\%, with weak two-dimensional density variations to allow departures from a straight column.  The imposed imbalance probes direction selection. An exactly reflection-symmetric
deterministic state remains in its symmetry sector and cannot test the growth of
a reflection-odd perturbation (Eqs.~\eqref{eq:kinetic-main} and
\eqref{eq:kinetic-torque-main}; SM Sec.~S.9.4).

We quantify the global heading order~\citep{MardiaJupp2000} by
\begin{equation}
P_m=\frac{\left|\int\dd^2\bm x\,\dd\phi\,
f(\bm x,\phi,t)\e^{\ii m\phi}\right|}
{\int\dd^2\bm x\,\dd\phi\,f(\bm x,\phi,t)}.
\label{eq:continuum-order-main}
\end{equation}
\(P_1\) measures net current in one direction, whereas \(P_2\) measures a shared
axis and remains large when equal populations move in opposite directions.

\begin{figure}[!t]
  \centering
  \includegraphics[width=\linewidth]{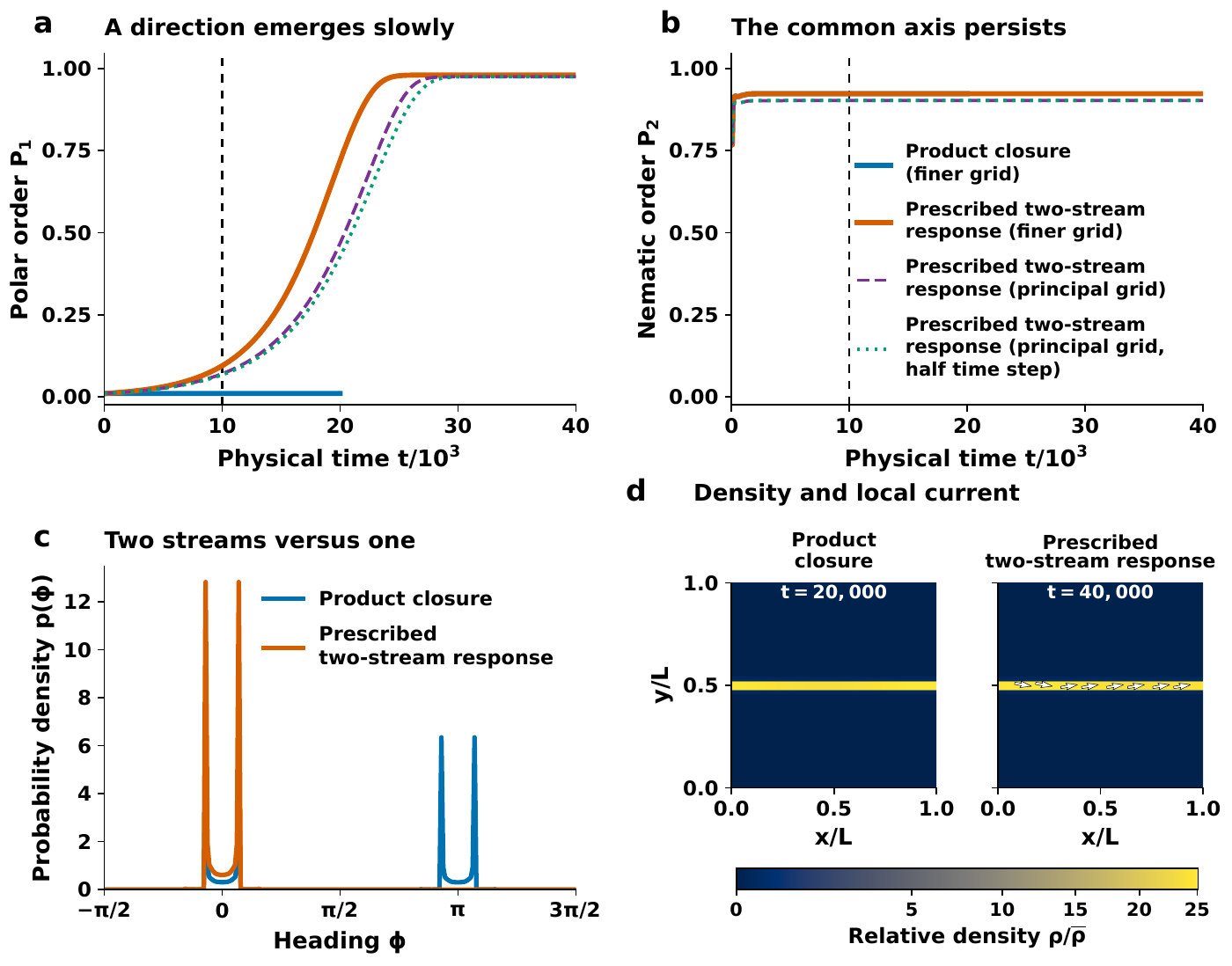}
  
  \caption{
  \textbf{Delayed polar flow in an attractive continuum column.}
  ``Product closure'' denotes the one-particle dynamics in
  Eqs.~\eqref{eq:kinetic-main} and \eqref{eq:kinetic-torque-main}, which treat
  neighboring agents as statistically independent. ``Prescribed two-stream response''
  denotes the two-stream population response in Eq.~\eqref{eq:pair-residence-main},
motivated by omitted pair information rather than derived from it, with
  prescribed rate \(a_{\rm pair}=0.035\) in rescaled time. All panels use
  distance-decaying attraction with \(\ell/L=1/16\).
  \textbf{(a,b)}, Global polar and nematic order
  (Eq.~\eqref{eq:continuum-order-main}): the prescribed response selects one direction
  while preserving the common axis; the product closure leaves nearly balanced
  counterpropagating populations. Time is in physical HT units; the vertical
  dashed line marks the end of the HT comparison interval. Solid curves use the
  finer grid; dashed and dotted curves show the two-stream response on the principal
  grid and with its time step halved, respectively. Grid sizes and time steps
  are given in SM Sec.~S.10.4.
  \textbf{(c)}, The normalized heading distribution (SM Eq.~(99)) retains two
  opposing directional sectors under product closure; one sector dominates with
  the prescribed two-stream response. The distributions are shown at
  \(t=20{,}000\) and \(40{,}000\), respectively, as in panel (d).
  \textbf{(d)}, Density divided by its spatial mean at the indicated times.
  Arrows show local mean velocity divided by \(v_0\): material flows along the
  column even when its density profile is nearly stationary. The prescribed
  response demonstrates direction selection under the assumed feedback. Its timing remains resolution-sensitive. SM Secs.~S.9 and S.10.4 give the
  construction and numerical controls, and SM Videos~3 and 4 show the dynamics.}
  
  \label{fig:ordered}
\end{figure}

For distance-decaying attraction at the HT range \(\ell/L=1/16\), the
product-closed continuum preserves a two-stream column for \(t=20{,}000\), or
40 box-crossing times.  Its polar order changes only from \(P_1=0.00957\) to
0.00975 (Fig.~\ref{fig:ordered}a), while nematic order grows from
\(P_2=0.840\) to 0.903 (Fig.~\ref{fig:ordered}b).  Refining the spatial and heading grids and halving the time step gives
\(P_1=0.00981\) and \(P_2=0.924\) (SM Sec.~S.10.4).  The two populations therefore develop a
sharper common axis without appreciably selecting one direction.
This persistent two-stream structure is consistent with HT's
bidirectional-column mechanism: the continuum retains that organization, but
does not reproduce the further direction selection reported by HT for
distance-decaying attraction~\citep{SalahshourCouzin2026HT}.

Adding the HT repulsive core, either to a constant attractive exterior or to
distance-dependent attraction with \(\ell/L=1/8\), also leaves the perturbed
column weakly polar: \(P_1=0.0100\) and 0.00987, respectively, at the same
final time.  Starting instead near angular disorder, or from two smooth
reconstructions of an HT initial particle configuration, gives modest polar
responses but no strongly polar product-closed state in these tests.
On the principal grid, trajectories through \(t=100{,}000\) preserve
this weakly polar response for all three radial laws (SM Sec.~S.10.4). These results describe the simulated interval. The sharp repulsive core requires finer spatial resolution (SM Secs.~S.9 and S.10.4).

An exact zero-width control helps separate existence from selection.
If all agents occupy distinct positions on a line parallel to a box axis
and point in the same direction along it, each neighbor is
exactly ahead or behind.  Every sine torque in Eq.~\eqref{eq:particle-law} then
vanishes separately, so the configuration translates rigidly for any radial
kernel.  The finite-width calculations determine whether the
continuum preserves such motion over a finite interval after the column is perturbed.
SM Sec.~S.9 gives the transverse linearization of the collinear state and the perturbed-column control.

\subsection{A prescribed two-stream response selects a polar current}
\label{sec:pair-selection}

HT suggests a physical reason to examine the pair information omitted by
Eq.~\eqref{eq:pair-factorization-methods}. Agents traveling together can remain
neighbors longer than agents passing in opposite directions.  With
distance-dependent attraction, these longer-lived encounters have greater
cumulative influence~\citep[SM Sec.~S.9.1.3]{SalahshourCouzin2026HT}; a repulsive
core also changes close encounters between the streams.  Heading-conditioned
neighborhood statistics beyond the product of one-particle fields contribute to
the pair-cue remainder \(\Delta G_n\) in the exact balance,
Eq.~\eqref{eq:pair-balance-methods} (SM Eqs.~(83) and (84)).  Pair factorization
discards this information even when the radial law retains distance decay or
collision avoidance.

A residence imbalance between the two counterpropagating populations could therefore reinforce one stream through its encounter history. Motivated by this possibility, we test a prescribed, symmetry-allowed two-stream response. Whether microscopic pair dynamics generate such a response remains open (SM Secs.~S.9.4 and S.9.5).

Once a nematic column has formed, its two dominant heading populations move in
opposite directions along a common axis. Let \(u\in[-1,1]\) be their local
population imbalance: \(u=0\) means equal populations, while \(u=\pm1\) means
that only one remains. We denote the signed global polarization along the
continuously tracked axis by \(P_\parallel\), and the strength of global
nematic order by \(P_2\) (Eq.~\eqref{eq:continuum-order-main}). To test whether
a small directional imbalance can grow, we use a minimal symmetry-preserving
normal form~\citep{Kuznetsov2004}:
\begin{equation}
  \begin{gathered}
  \left.\partial_tu\right|_{\rm pair}
  =a_{\rm pair}\,\mathcal A(P_2)P_\parallel(1-u^2),\\
  \mathcal A(P_2)=\operatorname{clip}\!\left(\frac{P_2-0.55}{0.45},0,1\right).
  \end{gathered}
  \label{eq:pair-residence-main}
\end{equation}
The activation \(\mathcal A(P_2)\) is zero for \(P_2\leq0.55\) and rises
linearly to one at \(P_2=1\). The clip operation keeps its value between zero
and one. The cutoff \(0.55\) is a phenomenological choice that restricts the
feedback to an already ordered column. The denominator \(0.45=1-0.55\)
normalizes the activation at perfect nematic order.

In Eq.~\eqref{eq:pair-residence-main}, \(a_{\rm pair}\) sets the response rate,
\(P_\parallel\) gives the sign of the net directional preference, and
\(1-u^2\) stops redistribution when one local population is exhausted.
Reversing all headings reverses the response, treating both directions equally;
at exact reflection symmetry, the response vanishes. During the pair substep,
\(P_\parallel\) and \(P_2\) are held fixed. Its exact integration preserves
the sum of each pair of opposite heading populations and keeps both
nonnegative. This population response is motivated by the pair information omitted from the one-particle equation. The remainder \(\Delta G_n\) itself has not been calculated (SM Sec.~S.9.5).

The prescribed response produces delayed direction selection.  With the prescribed rate \(a_{\rm pair}=0.035\) in rescaled time, the same initial
imbalance grows slowly at first, reaches \(P_1=0.468\) at \(t=20{,}000\), and
approaches \(P_1=0.975\) by \(t=40{,}000\).  Nematic order remains near
0.903. One stream takes over without destroying the common axis.  The final
heading distribution concentrates near one direction (Fig.~\ref{fig:ordered}c), and the
column carries a strong material current (Fig.~\ref{fig:ordered}d).  Material flows along the column even when its periodic density profile remains nearly stationary.

Strong polar flow persists when the time step, spatial grid, heading resolution,
and number of retained HT perceptual harmonics are varied. The selection time
and final heading profile remain resolution-sensitive (SM Sec.~S.10.4).
SM Videos~3 and 4 show polar flow with the prescribed response and the
product-closed bidirectional control using continuum fields only.
(SM Secs.~S.9, S.9.5, S.10.4, and S.11.)

The calculation shows that the prescribed reflection-odd response in
Eq.~\eqref{eq:pair-residence-main} can turn a long-lived two-stream column into
polar flow. Its rate, activation, and saturation are assumed. Deriving an effective
response from dynamical pair statistics and determining the polar state's
stability, basin, and size dependence remain open.  Pair-correlation-driven polarity
without explicit alignment also occurs in turn-away
hydrodynamics~\citep{DasCiarchiZhouYanZhangAlert2024}; the present mechanism
concerns selection between the two streams of an inhomogeneous HT column.
(SM Secs.~S.9.5, S.9.7, and S.10.4.)

\subsection{Pure repulsion: polar transport and transverse-band formation}
\label{sec:results-pure-repulsion}
\label{sec:methods-extended-regimes}

HT predicts that finite-range pure repulsion on a periodic square can organize
particles into bands that travel normal to their long axes
~\citep{SalahshourCouzin2026HT}.  We test the corresponding HT point
\(h=-1\), \(\sigma=1\), and \(L/\xi=4\) with the exact square-torus cue in
SM Eq.~(101).  Each continuum calculation starts from
a broadband, nearly isotropic field with no preformed band and evolves the
one-particle equations in Eqs.~\eqref{eq:kinetic-main} and
\eqref{eq:kinetic-torque-main}.  No pair-correlation term or heading-alignment
torque is present.

\begin{figure}[!t]
  \centering
  \includegraphics[width=\linewidth]{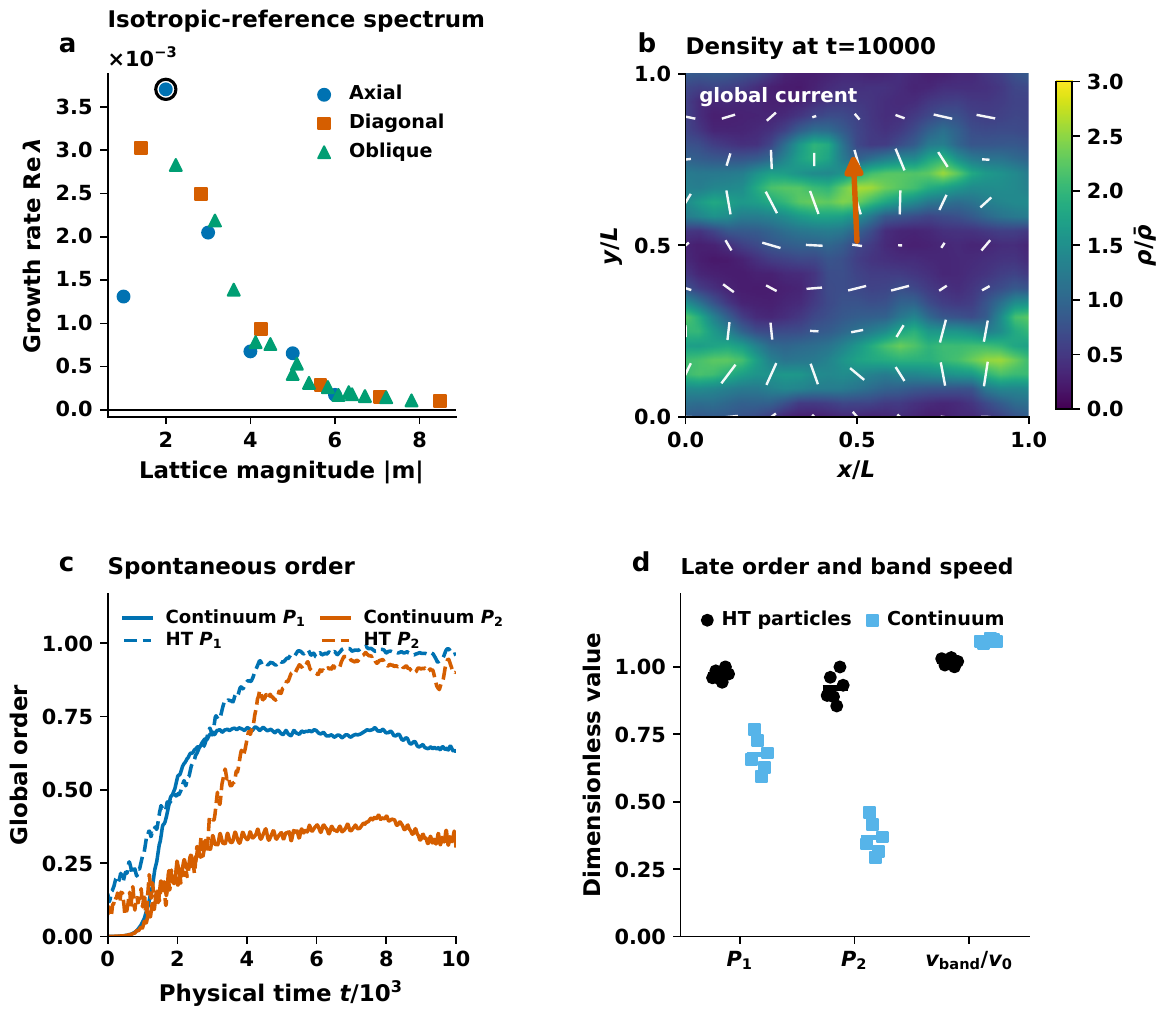}
  
  \caption{\textbf{Transverse-band motion in pure-repulsion continuum calculations.}
  All continuum calculations evolve
  Eqs.~\eqref{eq:kinetic-main} and \eqref{eq:kinetic-torque-main} with the
  square-torus cue in SM Eq.~(101), without a prescribed two-stream response. Physical
  parameters are the HT repulsion point \((h,\sigma,L/\xi)=(-1,1,4)\)
  ~\citep{SalahshourCouzin2026HT}.
  \textbf{(a)}, Growth rates estimated from the isotropic-reference
  approximation to Eq.~\eqref{eq:linear-operator-methods}, grouped by spatial
  wavevector direction. The horizontal axis measures wavevector magnitude in
  units of \(2\pi/L\). The axial two-band modes \((2,0)\) and \((0,2)\)
  grow fastest. The small uniform angular drift omitted in this estimate is
  retained in the nonlinear dynamics (SM Sec.~S.10.5).
  \textbf{(b)}, Density divided by its spatial mean at \(t=10{,}000\).
  White headless segments show local nematic directions; the orange arrow
  shows the global material-current direction, transverse to the bands. Its length is schematic.
  \textbf{(c)}, Evolution of global polar and nematic order
  (Eq.~\eqref{eq:continuum-order-main}). Curves are means over six independent
  realizations in each description: solid for the continuum and dashed for HT
  particles; blue for \(P_1\) and orange for \(P_2\).
  \textbf{(d)}, Polar order, nematic order, and band speed in the common
  late-time interval \(6{,}000\leq t\leq10{,}000\). Points represent independent realizations. Horizontal marks show their means. Band speed is measured from density-wave
  displacement (SM Eq.~(104)). SM Fig.~S16 shows this measurement directly.
  The continuum trajectories shown use Fourier spatial transport and first-order angular
  upwinding. SM Sec.~S.10.5 and Figs.~S15 and S17 compare numerical transport
  methods and show the sensitivity of late pattern selection.}
  
  \label{fig:pure-repulsion-continuum}
\end{figure}

An isotropic-reference approximation to
Eq.~\eqref{eq:linear-operator-methods}, using the square-domain cue in SM
Eq.~(101), estimates which spatial modes grow first.  It neglects the weak angular
drift produced by the square geometry at uniform density. The nonlinear calculations retain that drift (SM Sec.~S.10.5).  Among
the sampled lattice wavevectors, the leading instability is the axial two-band
mode \((m_x,m_y)=(2,0)\), together with its square-related partner \((0,2)\)
(Fig.~\ref{fig:pure-repulsion-continuum}a).  The growth rate and propagation frequency agree closely with direct
heading-grid calculations and with the early evolution of an initialized
eigenmode (SM Sec.~S.10.5).

With Fourier spatial transport, the axial two-band pattern develops spontaneously from broadband fluctuations. SM Fig.~S15a compares the early density-mode growth with the linear prediction. Agents move across the bands, along their common normal,
reproducing the distinctive transverse motion reported by HT
(Fig.~\ref{fig:pure-repulsion-continuum}b). Polar and nematic order develop
together as the bands form (Fig.~\ref{fig:pure-repulsion-continuum}c).
The comparison uses the same physical parameters and time interval in both
descriptions (Fig.~\ref{fig:pure-repulsion-continuum}d). Detailed trajectories,
band-displacement measurements, and numerical controls are given in
SM Sec.~S.10.5 and Figs.~S15 and S16. The continuum remains less globally
ordered than HT. Polar currents persist with conservative spatial and angular
transport, but the late density pattern changes (SM Sec.~S.10.5 and Fig.~S17).
A matched conservative grid and time-step comparison also retains a strong
current, while its onset and accumulated transport remain grid-sensitive
(SM Fig.~S18). Numerical transport, finite-particle effects, and pair correlations
can all contribute to the remaining particle--continuum differences. The
particular late morphology is implementation-sensitive.

SM Video~5 follows the band throughout the simulated interval, showing
how it forms and propagates across the periodic boundary.

The fitted phase speed measures density-wave propagation.  It can therefore slightly exceed \(v_0\) as agents leave the
rear of a band and join its front; by contrast, the center-of-mass identity in
SM Eq.~(79) remains
\(|\bm V_{\rm cm}|/v_0=P_1\leq1\).  Within this HT regime, the
one-particle
calculations support sustained polar transport under pure repulsion, without
a pair term or explicit heading alignment. Quantitative prediction of the
selected density pattern requires further numerical refinement. A distinct
alignment-free repulsion-assisted mechanism has been reported in turn-away
hydrodynamics~\citep{DasCiarchiZhouYanZhangAlert2024}.

This distinction between sharing an axis and selecting a direction
also survives a change in the decision profile. With a Gaussian bump matched
to the square bump in its first two perceptual coefficients, attraction
retains two-stream nematic order, whereas pure repulsion generates a
substantial polar current (SM Sec.~S.10.6 and Fig.~S19).
The comparison supports these forms of collective order without requiring
the sharp edges of the square bump.

\section{Discussion}

The central methodological contribution is a
\emph{representation-resolved} multiscale continuum theory downstream of HT.  The BBGKY
balance, pair factorization, Fokker--Planck equation, heading expansion, and
long-wave reduction are established tools~\citep{Risken1989,Spohn1991,Bertin2006,Peshkov2014}.  Here these operations preserve the microscopic perceptual information: HT's order-resolved signed spectrum \(q_n=c_n(\sigma)M_n(W)\) and the
radial law \(J(r)\) remain explicit in the nonlocal kinetic equation,
the finite-wavenumber operator, and the hydrodynamic transport coefficients.
Consequently, a macroscopic prediction can be traced to sensory acuity,
decision geometry, and interaction distance.
Inferring these ingredients separately requires independent measurements
or additional constraints, because several enter the collective coefficients
through products.  The exact BBGKY equation, the product-closed kinetic theory,
and the strict long-wave density hydrodynamics remain distinct levels of this
construction.  (SM Secs.~S.2--S.6.)

The comparison depends on the level of description.  Symmetry-based
polar hydrodynamics specifies the fields and allowed couplings after symmetry
breaking but does not derive them from a sensory representation
~\citep{TonerTu1995,Marchetti2013}.  Vicsek-derived kinetic theories connect fields
to microscopic rules that explicitly compare headings
~\citep{Bertin2006,Ihle2011,Peshkov2014}.  Scalar hydrodynamics of conventional
motility-induced phase separation requires no heading alignment, but its feedback
arises from collision-induced self-trapping or an effective loss of motility with
density~\citep{FilyMarchetti2012,RednerHaganBaskaran2013,CatesTailleur2015}.
A particularly relevant alignment-free hydrodynamic precedent for the present mechanism is Peruani's
vision-cone theory, which factorizes pairs, expands locally in density,
polarization, and nematic fields, and obtains aggregation, nematic bands, and moving
locally polar structures~\citep{Peruani2017}.  Selective attraction--repulsion has
also been coarse-grained without an explicit velocity-alignment rule
~\citep{GrossmannSchimanskyGeierRomanczuk2013}; pair-distribution theories describe
torque-driven phase separation~\citep{ZhangAlertYanWingreenGranick2021} and
repulsion-assisted polar motion~\citep{DasCiarchiZhouYanZhangAlert2024}.  A recent
nonlocal visual Boltzmann theory retains occlusion but uses a binary update that
averages headings~\citep{ItoUchida2025}.  The present theory retains the full order-resolved HT spectrum of a bearing-only interaction before gradient truncation, together with the exact pair remainder discarded by the one-particle closure.

At the level of results, density-directed reorientation and its
negative-diffusivity structure belong to an established class
~\citep{Peruani2017,ZhangAlertYanWingreenGranick2021}.  Eq.~\eqref{eq:threshold} connects that instability to HT's perceptual factors: \(c_1(\sigma)\) carries
sensory acuity, \(M_1(W)\) carries the directional decision, and \(\mu_2\)
carries the radial law.  It identifies macroscopic onset with HT's individual-level
direct-approach channel.  A second result is the conditional entry rule in
Eq.~\eqref{eq:harmonic-entry-methods}: HT supplies the indexed, signed channel
\(q_n\), whereas spatial symmetry and the required streaming path, derived here,
supply the exponent \(2n\).  A low-order local closure cannot recover this
order-by-order attribution without the same \(q_n\) inputs.  Finite-wavelength instabilities are likewise known in generic torque
field theories~\citep{SeseSansaLevisPagonabarraga2022}; what the present full
operator adds is a resolution of how their sign, scale, and heading-moment
composition depend on the separate HT coefficients and radial Hankel transforms.
(SM Secs.~S.5--S.8.)

Agreement is examined at three levels.  Exact
identities and operator reductions preserve the phases, signs, rotation covariance, and
reflection sectors of the heading-moment construction.  Nine independently discretized
kinetic-PDE cases then agree with the cutoff-converged finite-wavenumber operator across stable,
unstable, smooth-profile, alternative-kernel, and mixed-kernel regimes.  A direct
Monte Carlo finite-particle-generator calculation for the HT-derived particle law
resolves no projected discrepancy from the predicted initial heading-moment drift in either
a growing or decaying regime at this sample size.  This instantaneous comparison reaches the particle law before a
time step is taken; it neither re-examines the upstream neural derivation of \(q_n\)
nor assumes that initially absent pair correlations remain absent during later
evolution.  
(SM Sec.~S.10.)

The asymptotic and particle conclusions have a narrower scope.  The fourth-order
hydrodynamics is quantitative only in its measured small-\(k\) window, and a converged
heading-moment cutoff is essential because a low-order system can falsely suggest an
oscillatory onset.  Near threshold, the weakly nonlinear equation
in
Eq.~\eqref{eq:weak-nonlinear-methods}
\ reproduces the early growth of the fundamental and generated second spatial Fourier modes, but it does not determine
coexistence, transition order, or mature morphology.  In the many-neighbor
particle sequence, the growing sign is resolved at all four sizes and the decaying sign at
\(N=1{,}024\). Rate offsets and a near-neutral unresolved decay remain.  The fixed-range comparison is
limited by the finite-particle noise floor and therefore provides no evidence for
or against molecular chaos.

(SM Secs.~S.6, S.8, and~S.10.3.)

The long-time attractive-column calculations separate persistence of a common
axis from selection of a current along it.  With product closure, the perturbed
column remains nearly bidirectional through 200 box crossings on the principal grid.  This preserves
HT's distinction between a common axis and a selected direction.  A prescribed
reflection-odd term instead produces delayed polar flow while preserving the
nematic structure.  The late order is reproducible under time-step refinement
and exchange of the square axes, although the selection time is more sensitive.
The mechanism depends on the prescribed rate and saturating normal form. Its stability and thermodynamic scaling remain open.
Related alignment-free continuum structures and pair-induced polar effects are
discussed in Refs.~\citealp{Peruani2017,DasCiarchiZhouYanZhangAlert2024}.
(SM Secs.~S.9, S.9.5, and S.10.4.)

In the pure-repulsion regime, the one-particle nonlocal kinetic
continuum develops sustained polar currents from broadband nearly isotropic
fluctuations.    The bands in Fig.~6 share HT's transverse geometry and near-\(v_0\)
propagation, while their global polar and nematic orders remain lower over the
same time interval. Conservative spatial and angular transport retains a strong
current but changes the late density pattern (SM Sec.~S.10.5 and Fig.~S17).
Thus the calculations support collective transport more firmly than a specific
band geometry. Related alignment-free translation occurs in turn-away
hydrodynamics~\citep{DasCiarchiZhouYanZhangAlert2024}.

The present theory also identifies the next steps needed for a predictive
hydrodynamics of perceptual matter.  An effective two-stream response should be derived from the dynamical two-particle equation, including its rate and spatial dependence.  That derivation would allow the stability of
the polar column, the range of initial conditions from which it emerges, and its
size dependence to be predicted from microscopic parameters alone.  Describing
mature patterns will additionally require nonlinear theory beyond the near-onset
gradient expansion.  The ordered-state calculations also provide a practical
warning. An insufficient HT perceptual-harmonic cutoff can change the predicted
state qualitatively, so convergence must be demonstrated before assigning it
physical meaning.  Finally, because the continuum coefficients remain traceable to
sensory and neural parameters, a decisive empirical test would estimate those
parameters independently and predict continuum growth rates, wavelengths, and
collective motion without refitting.  

The finite-extent calculation treats neighbors as independently visible angular images. Occlusion and partial visibility would make a cue depend on intervening agents; elongated bodies would also couple apparent shape to their orientation. Extending the theory to these effects, and to physical exclusion between finite bodies, requires additional perceptual or pair information. The present geometric filter accounts only for angular extent.

The paper therefore makes four contributions.  First, it derives
continuum transport and instability from HT's sensory, decision, and
distance-dependent interaction parameters.  Second, it establishes the conditional
\(q_n\mapsto k^{2n}\) rule connecting each HT perceptual channel to the
spatial-derivative order at which it first affects conserved density.  Third, it identifies the pair information discarded by the one-particle BBGKY
closure and demonstrates direction selection by a prescribed, symmetry-allowed
two-stream response. Whether microscopic pair dynamics produce that response remains open.
Fourth, pure-repulsion continuum calculations sustain polar transport,
with transverse-band selection still sensitive to the transport discretization.

\noindent\begin{minipage}{\columnwidth}
\section*{Acknowledgments}

I thank Iain D. Couzin for fruitful discussions.

I acknowledge AI assistance (OpenAI ChatGPT 6 and 5.6, and Codex) in mathematical calculations, code development, numerical checks, plotting codes, manuscript review and consistency checks, writing and editing of the manuscript. I have verified all derivations, computations, {}and manuscript content and take full responsibility for the manuscript.
\end{minipage}

\section{Data and code availability}

Data,
trajectories, and code needed to reproduce the figures, tables, and videos will be
deposited in a permanent repository upon publication and are available from the
corresponding author.  SM Sec.~S.10 gives the numerical protocols.

\end{document}